%% file: java_performance_history.tex
\RequirePackage{fix-cm}
\RequirePackage{rotating}

\documentclass{svjour3}

\smartqed

\usepackage{algorithmic}
\usepackage{graphicx}
\usepackage{textcomp}
\usepackage{xcolor}
\usepackage{subfiles}
\usepackage{enumitem}
\usepackage{adjustbox}
\usepackage{float}
\usepackage{tabularx}
\usepackage{multirow}
\usepackage{caption}
\usepackage{amsmath,amsfonts}
\usepackage{url}
\usepackage{booktabs}
\usepackage{siunitx}
\usepackage{fancyvrb}
\usepackage{array}
\usepackage{collcell}
\usepackage[many]{tcolorbox}
\usepackage{todonotes}
\usepackage{makecell}
\usepackage{soul}
\usepackage{listings}
\usepackage{balance}
\usepackage{indentfirst}
\usepackage{wrapfig}
\usepackage{natbib}
\usepackage{afterpage}
\usepackage{pdflscape}

\newtcolorbox{boxK} {
    sharpish corners,
    boxrule = 0pt,
    toprule = 3pt,
    enhanced,
    fuzzy shadow = {0pt}{-2pt}{-0.5pt}{0.5pt}{black!35},
    boxsep=0mm,
    left=2mm,
    right=2mm,
    top=1.5mm,
    bottom=1.5mm
}

\definecolor{diffgreen}{rgb}{0.1,0.7,0.2}
\definecolor{diffred}{rgb}{0.8,0.1,0.1}

\lstdefinestyle{javastyle}{
    language=java,
    basicstyle=\ttfamily\footnotesize,
    breaklines=true,
    commentstyle=\color{gray},
    keywordstyle=\color{blue},
    stringstyle=\color{orange},
    showspaces=false,
    showstringspaces=false,
    showtabs=false,
    tabsize=2,
    frame=none,
    xleftmargin=0mm,
    moredelim=[is][\color{diffgreen}]{@g@}{@g@},
    moredelim=[is][\color{diffred}]{@r@}{@r@}
}

\newcommand{\customcell}[1]{%
  \begin{tabular}[t]{@{}l@{}r@{}}
    #1
  \end{tabular}%
}

\newcolumntype{C}{>{\collectcell\customcell}c<{\endcollectcell}}

\newcommand\numProjects{15}

\newcommand\numBenchmarkedCommits{739}
\newcommand\numMethodChanges{1,499}
\newcommand\numSignificantMethodChanges{479}
\newcommand\numLabeledMethodChanges{958}

\newcommand\jibURL{https://github.com/kavehshahedi/java-instrumentation-buddy}
\newcommand\jpbURL{https://github.com/kavehshahedi/java-parser-buddy}

\begin{document}

\title{An Empirical Study on Method-Level Performance Evolution in Open-Source Java Projects}

\titlerunning{Method-Level Performance Evolution in Java Projects}

\author{Kaveh Shahedi \and
        Nana Gyambrah \and
        Heng Li \and
        Maxime Lamothe \and
        Foutse Khomh}

\authorrunning{Shahedi et al.}

\institute{Kaveh Shahedi, Nana Gyambrah, Heng Li, Maxime Lamothe, and Foutse Khomh are with the Department of Computer Engineering and Software Engineering at Polytechnique Montréal, Canada \\
              Heng Li is the corresponding author: \email{heng.li@polymtl.ca}
}

\date{Received: date / Accepted: date}

\maketitle

\begin{abstract}
\subfile{sections/abstract}

\end{abstract}

\section{Introduction}
\label{sec:introduction}
\subfile{sections/introduction}

\section{Background}
\label{sec:background}
\subfile{sections/background}

\section{Related Works}
\label{sec:related-works}
\subfile{sections/related_works}

\section{Methodology}
\label{sec:methodology}
\subfile{sections/methodology}

\section{Results}
\label{sec:evaluations}
\subfile{sections/evaluations}

\section{Discussion}
\label{sec:discussion}
\subfile{sections/discussion}

\section{Threats to Validity}
\label{sec:threats}
\subfile{sections/threats}

\section{Conclusion}
\label{sec:conclusion}
\subfile{sections/conclusion}

\bibliographystyle{spbasic}
\bibliography{java_performance_history}

\end{document}

%% file: sections/abstract.tex
Performance is a critical quality attribute in software development, yet the impact of method-level code changes on performance evolution remains poorly understood. While developers often make intuitive assumptions about which types of modifications are likely to cause performance regressions or improvements, these beliefs lack empirical validation at a fine-grained level. We conducted a large-scale empirical study analyzing performance evolution in {\numProjects} mature open-source Java projects hosted on GitHub. Our analysis encompassed {\numBenchmarkedCommits} commits containing {\numMethodChanges} method-level code changes, using Java Microbenchmark Harness (JMH) for precise performance measurement and rigorous statistical analysis to quantify both the significance and magnitude of performance variations. We employed bytecode instrumentation to capture method-specific execution metrics and systematically analyzed four key aspects: temporal performance patterns, code change type correlations, developer and complexity factors, and domain-size interactions. Our findings reveal that 32.7\% of method-level changes result in measurable performance impacts, with regressions occurring 1.3 times more frequently than improvements (18.5\% vs 14.2\%). Contrary to conventional wisdom, we found no significant differences in performance impact distributions across code change categories, challenging risk-stratified development strategies. Algorithmic changes demonstrate the highest improvement potential (25.6\%) but carry substantial regression risk (33.9\%). Senior developers produce more stable changes with fewer extreme variations, while code complexity correlates with increased regression likelihood (15.5\% to 23.6\% across complexity levels). Domain-size interactions reveal significant patterns, with web server + small projects exhibiting the highest performance instability (42.2\%). Our study provides empirical evidence for integrating automated performance testing into continuous integration pipelines and contributes a comprehensive dataset for future performance analysis research.

%% file: sections/introduction.tex
In the realm of software development, performance is a critical factor that directly impacts system efficiency and, consequently, user experience~\citep{smith2002performance, smith2007introduction,bondi2015foundations, forstandardizationiso2005isoiec}. As applications evolve through their development lifecycle, they undergo various code changes that can either enhance or degrade their performance~\citep{foo2015industrial,chen2022adaptive,nguyen2012automated,shang2015automated,ahmed2016studying,liao2021locating}. This is natural since, initially, developers often prioritize functionality to deliver working software rapidly, sometimes at the expense of sacrificing performance~\citep{tuma2018threat, yamashita2013developers}. As software matures and moves closer to production environments, especially those with limited resources, the focus shifts toward optimizing performance to ensure efficient and optimal execution~\citep{jin2012understanding}.

Every software system is getting ever more complex, with interconnected components that span from basic code blocks to sophisticated services~\citep{lehman1980programs, wang2021promises}. Among the most performance-influential components are application methods (i.e., functions), which form the core logic and functionality of the program. Existing practices often consider the method level as a natural measurement unit, with examples such as performance profilers like JProfiler~\citep{jprofiler} and tracers such as OpenTelemetry~\citep{opentelemetry}. Each method is designed to perform a specific responsibility and seamlessly integrate with others to accomplish a final task. Consequently, newly introduced changes to these methods can lead to performance variations, which developers must carefully evaluate throughout the development lifecycle. Notably, micro-benchmarks have been widely used to measure method-level performance, providing granular insights into execution efficiency~\citep{laaber2018evaluation, georges2007statistically}.

On this account, performance fluctuations (i.e., either improvements or regressions) are common~\citep{alcocer2015tracking, chen2017exploratory} throughout the software lifecycle. Even code changes that do not alter functionality can introduce unintended slowdowns, bottlenecks, or system crashes. Such fluctuations, ranging from significant degradations to notable enhancements, are common across almost all applications~\citep{jin2012understanding, zaman2012qualitative, selakovic2016performance}. Understanding how these performance changes occur over time is essential for developers who are aiming to maintain and improve software efficiency.

Despite the importance of performance evolution, there is a lack of comprehensive studies that systematically analyze performance changes at a fine-grained level across different projects, particularly regarding method-level code changes. Prior research has largely focused on specific performance optimization techniques~\citep{georges2007statistically, laaber2018evaluation, chen2016robust}, performance bug detection~\citep{nistor2013toddler, han2018perflearner}, or case studies of individual systems~\citep{bezemer2015understanding}. This leaves a gap in understanding the broader patterns and trends of performance evolution in software projects, especially at the method level, where many critical performance-related changes occur. Also, most of the previous works concentrated on analyzing the performance fluctuation at a more abstract level of the software development lifecycle (i.e., release level)~\citep{ferme2017towards, chen2020perfjit}, leading to ignoring performance fluctuations that come from minor updates (i.e., commit level).

Focusing on method-level performance is essential for understanding the nuances of performance fluctuations in software applications. Since methods are the fundamental units of programs, alterations at this level can cause significant variations in performance. By analyzing how changes to method-level code impact their performance, we obtain precise insights into the effects of specific modifications on the application's efficiency. This fine-grained approach enables developers to accurately identify the exact locations within the codebase where performance regressions or improvements occur. Thus, targeted optimizations become feasible, leading to enhanced overall software quality.

To further emphasize the importance of analyzing performance-related aspects of method-level code changes, Figures~\ref{fig:code-improv-example} and~\ref{fig:code-reg-example} illustrate how simple code changes can significantly impact performance, either as improvements or regressions. In Figure~\ref{fig:code-improv-example}, a minor adjustment in the approach to performing a specific task (e.g., switching from Java's String concatenation to using \textit{StringBuilder}) resulted in an over $96\%$ decrease in execution time. Conversely, Figure~\ref{fig:code-reg-example} highlights how unintended performance-related code changes can introduce considerable regressions (i.e., $70\%$ slowdown), influenced by factors such as Java's autoboxing mechanism for primitive types or changes in Just-In-Time (JIT) compiler optimizations in the Java Virtual Machine (JVM) environment.

\begin{figure}[]
\begin{minipage}{0.48\textwidth}
\begin{lstlisting}[style=javastyle]
// Version 1
public String builderV1(List<String> parts) {
    String result = "";
    for (String part : parts)
        result += part;
    return result;
}
\end{lstlisting}
\end{minipage}
\hfill
\begin{minipage}{0.48\textwidth}
\begin{lstlisting}[style=javastyle]
// Version 2
public String builderV2(List<String> parts) {
    @g@StringBuilder result = new StringBuilder();@g@
    for (String part : parts)
        @g@result.append(part);@g@  
    return result.toString();
}
\end{lstlisting}
\end{minipage}
\caption{There are two approaches for concatenating strings from a list. The second approach (V2) is enhanced by using a \textit{StringBuilder} instead of direct string concatenation. When averaged over $100$ trials with a list containing $100,000$ strings, each ranging from $5$ to $9$ characters in length, the first approach (V1) required $1.27$ seconds to complete. In contrast, the optimized V2 only took $4.5$ milliseconds, resulting in a $96\%$ increase in speed. Additionally, as the size of the input grows, the performance gap widens quadratically because V1 continuously creates new string objects and incurs overhead from garbage collection.}
\label{fig:code-improv-example}
\end{figure}

\begin{figure}[]
\begin{minipage}{0.44\textwidth}
\begin{lstlisting}[style=javastyle]
// Version 1
public double processorV1(Order order) {
    return order.getItems()
        .stream()
        .mapToDouble(o -> 
            o.getPrice() * o.getQuantity())
        .sum();
}
\end{lstlisting}
\end{minipage}
\hfill
\begin{minipage}{0.52\textwidth}
\begin{lstlisting}[style=javastyle]
// Version 2
public double processorV2(Order order) {
    return @g@Optional.ofNullable(order)
        .map(Order::getItems)
        .orElse(Collections.emptyList())@g@
        .stream()
        @g@.filter(Objects::nonNull)
        .filter(item -> 
            Optional.ofNullable(item.getPrice())
                .isPresent())@g@
        .mapToDouble(item ->
            @g@Optional.ofNullable(item.getPrice())
                .orElse(0.0)@g@ * item.getQuantity())
        .sum();
}
\end{lstlisting}
\end{minipage}
\caption{Two variants of a method for calculating order totals demonstrate the impact of defensive programming on performance. Version 2 (right) was adjusted to avoid null pointer exceptions by using \textit{Optional} wrappers around values, but this added safety resulted in a considerable performance decline. In tests involving $100,000$ orders, each with $100$ items, and averaged over $100$ runs, version 1 (left) finished in $50$ milliseconds, whereas the newer version took $84$ milliseconds, making it approximately $70\%$ slower. The significant slowdown is attributed to V2's creation of numerous optional objects and unnecessary null checks, which increase garbage collection overhead and hinder JVM optimizations that the more straightforward V1 implementation could achieve.}
\label{fig:code-reg-example}
\end{figure}

To address this gap, we conducted a large-scale empirical study analyzing the performance evolution of {\numProjects} mature and well-developed open-source Java projects hosted on GitHub. Our study focuses on how the performance of these projects evolves over time due to the method-level code changes introduced at the commit level. By investigating the git commit history of each project, we identified code changes made to Java files in each commit, specifically those changes within the scope of the programs' methods. We then compared each commit to its predecessor to assess the performance impact of the newly introduced changes.

To identify relevant commits and methods for our analysis, we first filtered each project's commit history for changes to Java files containing JMH benchmarks. For each qualifying commit, we extracted only those methods that were directly modified. This targeted approach explains why we analyzed {\numBenchmarkedCommits} commits and collected data on {\numMethodChanges} method-level changes across the studied projects, which is a subset of all commits in these mature projects. To evaluate the performance differences between consecutive versions, we used the dedicated Java Microbenchmark Harness (JMH) module of each project to execute benchmarks targeting the modified methods. While JMH provides overall execution time measurements, we needed finer-grained insights into method behavior. Therefore, we additionally employed a lightweight Java instrumentation agent attached to the Java Virtual Machine (JVM) to instrument these methods during benchmarking. This instrumentation was essential to capture method-specific execution details that JMH alone cannot provide, such as internal method state changes and precise execution paths. This approach provided detailed insights into the performance behavior of the modified methods, capturing metrics such as minimum, maximum, and average execution times when executing the benchmarks~\citep{shahedi2025jperfevo}.

By aggregating the performance data obtained from microbenchmarking the studied programs, we performed various statistical analyses to identify patterns of performance improvements or degradations throughout their development lifecycle. This allowed us to examine how performance evolves over time. We also analyzed the relationship between the type of code changes (e.g., algorithmic, data structure) and their impact on performance. Additionally, we correlated the developers' experience with the performance changes resulting from their code modifications, enabling us to assess whether a noticeable and significant pattern exists between drastic performance fluctuations and the experience of the committer. Finally, we examined whether significant differences in performance evolution exist between projects from different domains, as well as those varying in size and complexity.

\noindent Our study aims to answer the following research questions:
\begin{itemize}
    \item \textbf{RQ1}: \textit{What are the patterns of performance changes in Java projects over time?} We seek to identify and analyze trends in performance improvements or degradations across the studied projects, providing insights into the overall trajectory of performance evolution over time. 
    \item \textbf{RQ2}: \textit{How do different types of code changes correlate with performance impacts?} We aim to understand the relationship between specific types of code modifications (e.g., algorithmic or data structure) and their effects on performance, identifying which changes are most likely to impact performance positively or negatively.
    \item \textbf{RQ3:} \textit{How do developer experience and code change complexity relate to performance impact magnitude?} We seek to identify the correlation between developmental factors, such as developers' experience and complexity of the code changes, and their impacts on performance through code changes.
    \item \textbf{RQ4}: \textit{Are there significant differences in performance evolution patterns across different domains or project sizes?} We investigate whether domain-specific or size-related trends exist in performance evolution, offering insights into how project characteristics influence performance patterns.
\end{itemize}

Through this research, we make the following contributions:
\begin{enumerate}
    \item \textit{Comprehensive Analysis of Performance Evolution Introduced by Method-Level Code Changes}: We provide an extensive examination of performance evolution over time across multiple mature Java projects, highlighting common patterns and trends.
    \item \textit{Correlation Between Code Change Types and Performance Impacts}: We develop a taxonomy of method-level code changes and analyze how different types of modifications correlate with performance improvements or regressions.
    \item \textit{Contribution of Developmental Factors to Performance Variations:} We examine how some factors like developer expertise and code complexity influence performance outcomes, providing actionable insights to manage regressions, and also, plan improvements.
    \item \textit{Insights into Domain and Size-related Performance Patterns}: We explore how performance evolution differs across various domains and project sizes, contributing to a deeper understanding of performance dynamics in software development.
    \item \textit{Dataset for Future Research}: We share a dataset comprising {\numMethodChanges} method-level code changes' performance assessment, serving as a valuable resource for subsequent studies on performance analysis and optimization.
\end{enumerate}

We share a replication package to encourage future work to replicate or build on our work. All the tools, scripts, analysis notebooks, and other materials that were used in this study are provided in this replication package.\footnote{\url{https://github.com/mooselab/empirical-java-performance-evolution}}.

The remainder of this paper is organized as follows. First, Section~\ref{sec:background} covers some background and terminologies used in this study. Section~\ref{sec:related-works} discusses related work on performance analysis and evolution in software systems. Section~\ref{sec:methodology} describes our methodology, including data collection, benchmarking procedures, and analytical techniques. In Section~\ref{sec:evaluations}, we present the results of our empirical study, addressing each research question in detail. Sections~\ref{sec:discussion} and~\ref{sec:threats} discuss the implications of our findings and potential threats. Finally, section~\ref{sec:conclusion} concludes the paper and outlines directions for future research.

%% file: sections/background.tex
This study explores several key areas of software engineering, including the microbenchmarking of Java applications, the evaluation of performance variations in method-level code changes, and code instrumentation. In this section, we present a comprehensive overview of each domain.

\subsection{Microbenchmarking of Java Applications}
Benchmarking is a fundamental aspect of software engineering and performance engineering, focusing on analyzing various performance-related aspects of software systems. Developers commonly use benchmarking to assess the efficiency of their products, either through synthesized use cases or real-world data. Additionally, it serves as a comparative tool, enabling developers to evaluate performance variations between different versions of their applications. This helps identify any unexpected anomalies introduced during development. One of the key advantages of benchmarking is its flexibility, it is not restricted to the final stages of development. Instead, it can be implemented throughout the development lifecycle, allowing developers to actively monitor and assess the efficiency of their products in real time.

Microbenchmarking is a specialized subset of benchmarking that focuses on evaluating the performance of isolated sections of a program, often referred to as micro-operations. While general benchmarking assesses the overall performance of an entire software system, microbenchmarking operates at a more granular level, enabling developers to analyze the performance characteristics of individual components in an isolated environment. This approach allows for precise assessment of newly introduced code changes, whether at the block, method, class, or higher levels of the program. It is important to address why traditional unit tests are not used for analyzing the performance of fine-grained regions of applications, as opposed to dedicated microbenchmarks. Unit tests are primarily designed to evaluate the functionality of application components, rather than their performance characteristics. Consequently, relying on unit tests for performance analysis would lead to inaccurate and unreliable results due to environmental influences and insufficient observations and executions. Therefore, in this study, we opted for microbenchmarks, as they provide a more precise and robust means of evaluating performance changes between different versions of programs.

In the realm of Java applications, performance microbenchmarking plays a crucial role. In addition to the expected performance variations introduced by new code changes, the Java Virtual Machine (JVM) significantly influences performance characteristics. Features such as Just-in-Time (JIT) compilation, Garbage Collection (GC), and various runtime code optimizations are common phenomena associated with the JVM. Furthermore, Java's extensive feature set, including advanced object, oriented capabilities, diverse libraries, and robust frameworks, makes performance microbenchmarking essential. By applying microbenchmarking techniques, developers can mitigate the risk of encountering unexpected performance regressions in production environments.

To facilitate the implementation of microbenchmarking in Java applications, OpenJDK offers the Java Microbenchmark Harness (JMH). JMH is specifically designed to streamline the process of designing, executing, and analyzing microbenchmarks in Java programs. Its comprehensive features,such as warm-ups, customizable iteration counts, and support for multiple forks (e.g., multiple JVM invocations), significantly reduce the error rate caused by environmental factors. However, despite its powerful and versatile features, JMH does have some drawbacks that can make it challenging for developers to fully leverage its capabilities for performance analysis. One significant challenge is the potential for non-deterministic results, which can complicate the interpretation of benchmarking outcomes. However, as we did in this study, this issue can often be mitigated by increasing the number of iterations and forks, ensuring more reliable and statistically significant results.

\subsection{Method-Level Code Change Performance Analysis}
Methods are the fundamental building blocks of applications, each serving specific responsibilities to deliver the final results to the user. When developers introduce new code changes to methods, their primary intent might be to modify functionality rather than performance. However, such changes can inadvertently lead to unexpected performance variations. Analyzing method-level code changes is therefore essential for gaining a comprehensive understanding of how individual components of an application evolve over time (e.g., through commits). This analysis also helps developers understand how different types of code changes affect method performance (e.g., algorithmic or data structure), enabling them to better prevent unintended performance regressions.

To accurately and effectively evaluate the performance evolution introduced by code changes at the method level in Java applications, microbenchmarking, specifically using JMH benchmarks in Java applications, provides a robust solution. With the ability to narrow down to method-level granularity, JMH allows developers to apply benchmark annotations to performance-critical methods in their applications. In this study, we aim to analyze the performance evolution between two versions of a method, following a code change, by employing JMH-based microbenchmarking. This approach enables precise detection of performance changes, whether they represent improvements, degradations, or remain neutral.

\subsection{Software Instrumentation}
Software instrumentation, also known as tracing, is a more advanced form of logging in which a separate component is integrated into the system to collect and store various types of information as instructed. Instrumentation enables developers and system administrators to gain a detailed understanding of the flow of actions and requests within the system, from initiation to completion. Unlike traditional logging, where developers manually insert logging statements into their programs, instrumentation is often automated or requires minimal manual intervention. Furthermore, instrumentation can be applied at different levels, ranging from the kernel to the application level. It is important to note that the granularity of instrumentation can significantly impact system performance. As such, developers must carefully optimize their instrumentation configurations to minimize overhead while ensuring the required level of detail is captured.

In this study, we utilized application-level instrumentation to record the executions of modified methods in each commit, enabling us to observe and evaluate performance variations across different versions. By attaching a Java instrumentation agent to the JVM running the JMH microbenchmarks, we were able to capture method executions effectively.

%% file: sections/related_works.tex
This section categorizes related works into three main areas: performance regression analysis and detection, the development of tools and automated approaches for performance benchmarking, and the analysis of performance anti-patterns.

\subsection{Performance Regression Analysis and Detection}

Several studies have explored the nature and causes of performance regressions, aiming to understand how code changes impact software performance~\citep{jin2012understanding, zaman2012qualitative, nguyen2012automated, selakovic2016performance, ding2020towards, heger2013automated, malik2013automatic, shahedi2023tracing}. Chen et al. conducted an exploratory study focusing on performance regressions in the evolution of projects like Hadoop, highlighting that a significant portion of regressions arises from changes intended to fix other issues~\citep{chen2017exploratory}. Their study underscores the need to monitor code changes during development to avoid unintended performance degradation. While Chen et al. focused on project-level performance regressions, our work provides a more granular analysis by examining how specific method-level changes contribute to performance variations.

Luo et al. proposed the PerfImpact system, which utilizes genetic algorithms and change impact analysis techniques to detect performance regressions~\citep{luo2016mining}. Their approach mines execution traces and evaluates the influence of code changes, which aligns with our methodology of analyzing method-level changes. However, our approach differs by emphasizing benchmarking directly at the method level, providing more precise insights into performance changes as they occur.

Nguyen et al., in an industrial case study, demonstrated the utility of mining past performance regression data to predict future regressions~\citep{nguyen2014industrial}. By leveraging historical performance data, their approach achieves up to 80\% accuracy in identifying regression causes. While their work emphasizes the use of historical data, our analyses focus on investigating the characteristics that contribute to significant performance variations.

Liao et al.~\citep{liao2021locating} present an automated approach for locating the root causes of performance regressions using field operational data instead of resource-intensive in-house performance testing. Their methodology employs black-box performance models and statistical techniques to analyze runtime activities and identify performance bottlenecks. Evaluations across three open-source projects and an industrial system demonstrate the effectiveness of this approach in detecting both synthetic and real-world performance regressions. While Liao et al. leverage field operational data, our approach focuses on microbenchmarking method-level code changes. Since these changes are generally trivial between versions, our microbenchmarking process remains feasible even in resource-constrained environments, providing developers with targeted insights into specific code components that affect performance.

Shang et al.~\citep{shang2015automated} present an automated approach for detecting performance regressions by analyzing collected performance counters. Their method clusters counters to minimize the number of required models and uses statistical tests to select target counters for regression modeling. Applied to two case studies, their approach successfully detects both synthetic and real-world performance regressions, outperforming traditional methods. Unlike Shang et al., who focus on system-level performance counters, our approach targets method-level code changes, enabling more precise localization of performance issues directly tied to specific code modifications.

Foo et al.~\citep{foo2015industrial} address the challenge of performance regression detection in heterogeneous test environments. Traditional approaches assume consistent test conditions, limiting their real-world applicability. They propose an ensemble-based method to model system behavior using prior test runs across different environments. Deviations in new tests are flagged and aggregated to distinguish true regressions from environment-specific variations. Case studies on open-source and enterprise systems demonstrate that their proposed approach outperforms state-of-the-art methods, with stacking proving more effective than bagging. While they address the challenge of environmental heterogeneity in performance testing, our approach complements this by focusing on the consistency of method-level performance analysis, which can be integrated into their framework to enhance regression detection in varied environments.

Chen et al. integrated performance monitoring within CI systems to provide developers with real-time insights~\citep{chen2019analyzing}. This aligns with our approach of tracking performance evolution across commits; however, our work incorporates a finer granularity by analyzing method-level changes through benchmarking.

In summary, while these studies contribute significantly to understanding performance regressions, our work emphasizes analyzing the rationales and characteristics that lead to performance regressions through method-level changes.

\subsection{Tools and Approaches for Performance Benchmarking}

Several works have been conducted to analyze the process of detecting and analyzing performance regressions~\citep{georges2007statistically, chen2016robust, laaber2018evaluation, sandoval2016learning, alcocer2019performance}. PeASS, introduced by Reichelt et al., identifies performance changes by transforming existing unit tests into performance benchmarks~\citep{reichelt2019peass}. Their tool automates regression testing, reducing the manual effort required to define benchmarks. This approach supports our work, as our investigation leverages an automated performance testing approach for identifying performance variations. However, while PeASS transforms existing unit tests, our approach directly targets method-level code changes, providing a more focused analysis of performance impacts across software versions.

The study by Traini et al. on the impact of refactoring operations reveals that even minor refactoring can significantly affect performance~\citep{traini2021software}. They demonstrate that developers must be cautious when modifying performance-critical code components. This complements our work, which integrates benchmarking directly into the development cycle to detect such changes early.

Georges et al.~\citep{georges2007statistically} introduced a comprehensive statistical approach for evaluating Java performance. They developed a framework that mitigates the inherent non-determinism of the JVM by employing multiple benchmark runs and applying statistical tests. Their study revealed that prevalent Java performance evaluation methods were not always precise, and they formulated guidelines for conducting statistically robust performance assessments. This work aligns with our microbenchmarking procedure to ensure robust results. We extend Georges et al.'s statistical foundation by applying these principles specifically to method-level changes, enabling more targeted performance analysis within the software development lifecycle.

Alcocer et al.~\citep{sandoval2016learning} analyze the impact of source code changes on performance across 1,288 versions from 17 open-source projects. They identify 10 types of performance-affecting changes and propose a cost model using horizontal profiling to predict regressions efficiently, detecting 83\% of regressions over 5\% and all above 50\%. In a related study~\citep{alcocer2019performance}, the authors introduce the Performance Evolution Matrix, an interactive visualization combining time-series comparisons and execution graphs to track performance changes. A controlled experiment shows that this method reduces the effort required to detect regressions while maintaining accuracy compared to traditional profiling tools. These works follow a similar approach to our data collection procedure through microbenchmarking, while our work is more fine-grained.

In conclusion, while tools like PeASS and approaches from Traini et al. provide valuable insights into performance regressions, our work differentiates itself by implementing a continuous and automated pipeline focused on benchmarking at the method level to retrieve performance insights.

\subsection{Performance Anti-Patterns}

Performance anti-patterns have been widely studied as a means to identify recurring design and implementation mistakes that negatively affect software performance~\citep{smith2000antipatterns, smith2002performance, smith2002new, parsons2004framework, xu2023mining, harrison1999design, taba2013predicting, chen2017characterizing, lui2021real}. These anti-patterns provide a structured way to recognize inefficiencies and offer refactoring solutions.

Smith and Williams pioneered the study of performance anti-patterns, documenting common pitfalls in software systems. Their foundational work~\citep{smith2000antipatterns} introduced a catalog of performance anti-patterns and outlined strategies to mitigate their impact. Later, they expanded this catalog by adding three additional anti-patterns, highlighting new ways developers inadvertently introduce performance issues~\citep{smith2002new}.

Parsons and Murphy~\citep{parsons2004framework} proposed a framework for the automatic detection of performance anti-patterns in component-based software. Their approach leverages runtime analysis to monitor system behavior, detect anti-pattern occurrences, and visualize their effects on performance. This framework aligns with our focus on automated performance analysis, although our approach emphasizes method-level benchmarking to detect performance anomalies.

Xu et al.~\citep{xu2023mining} explored the relationship between object-relational mapping (ORM) performance anti-patterns and code clones. Through empirical analysis across multiple open-source ORM systems, they identified four recurring anti-patterns linked to redundant ORM code and proposed strategies to mitigate their effects. Their study emphasizes the significance of context-aware code reuse, which complements our focus on identifying inefficiencies introduced by method-level changes. While Xu et al. concentrate specifically on ORM-related anti-patterns, our work provides a more general framework for detecting performance issues across various types of code modifications at the method level.

Harrison et al.~\citep{harrison1999design} examined the impact of object-oriented design patterns on software quality attributes, including maintainability and performance. Their findings suggest that while design patterns can enhance software quality, certain patterns may introduce performance trade-offs. This aligns with our approach to analyzing the performance impact of software design decisions at a finer granularity. Where Harrison et al. examine design patterns broadly, our method provides empirical performance data at the method level, offering more precise insights into how specific implementations of design patterns affect performance in practice.

Taba et al.~\citep{taba2013predicting} examine how antipatterns can improve bug prediction models, addressing the limitations of traditional software metrics. The authors analyze multiple software versions to assess the relationship between antipatterns and bug density, proposing new metrics based on antipattern history. Their findings show that antipattern-based metrics enhance bug prediction accuracy, outperforming traditional approaches. Our work can contribute as a fundamental resource for this type of study, as it can provide performance anti-patterns across different program versions, which can improve the accuracy of prediction models.

Chen et al.~\citep{chen2017characterizing} investigate the challenge of maintaining high-quality logging code due to the absence of well-defined guidelines. By analyzing the development history of three open-source systems, the authors identify six logging anti-patterns and develop LCAnalyzer, a static analysis tool to detect them. Case studies demonstrate that LCAnalyzer achieves high recall and satisfactory precision, successfully identifying anti-patterns in multiple projects. The findings highlight the importance of structured logging practices, with positive feedback from developers supporting the tool's usefulness. As our work also extracts code changes between versions, correlating the performance impact of logging code modifications can be investigated precisely.

Lui et al.~\citep{lui2021real} propose a real-time detection approach for the "More is Less" software performance anti-pattern in MySQL databases using dynamic analysis and machine learning. A support vector machine is trained on system metrics, database metrics, and MySQL status variables, and then integrated into a runtime analysis tool for real-time classification. The results demonstrate high accuracy, achieving 99.1\% sensitivity in detecting the anti-pattern. By incorporating anti-pattern detection techniques similar to their work but using the outputs of our method-level performance analysis, developers can train more robust and accurate performance models to quickly pinpoint performance bugs across diverse software components, not limited to database operations.

In summary, while previous research has significantly contributed to identifying and mitigating performance anti-patterns, our work emphasizes integrating these insights into an automated benchmarking pipeline to systematically detect and address performance inefficiencies at the method level.

%% file: sections/methodology.tex
We aim to analyze the performance evolution of Java programs by investigating method-level changes during their development history. Our methodology involves selecting suitable projects, processing commits, building and executing performance benchmarks, and analyzing the performance impacts (i.e., improvement or degradation) of code changes. This section provides a detailed explanation of each step.

\begin{figure*}
    \centering
    \includegraphics[width=\textwidth]{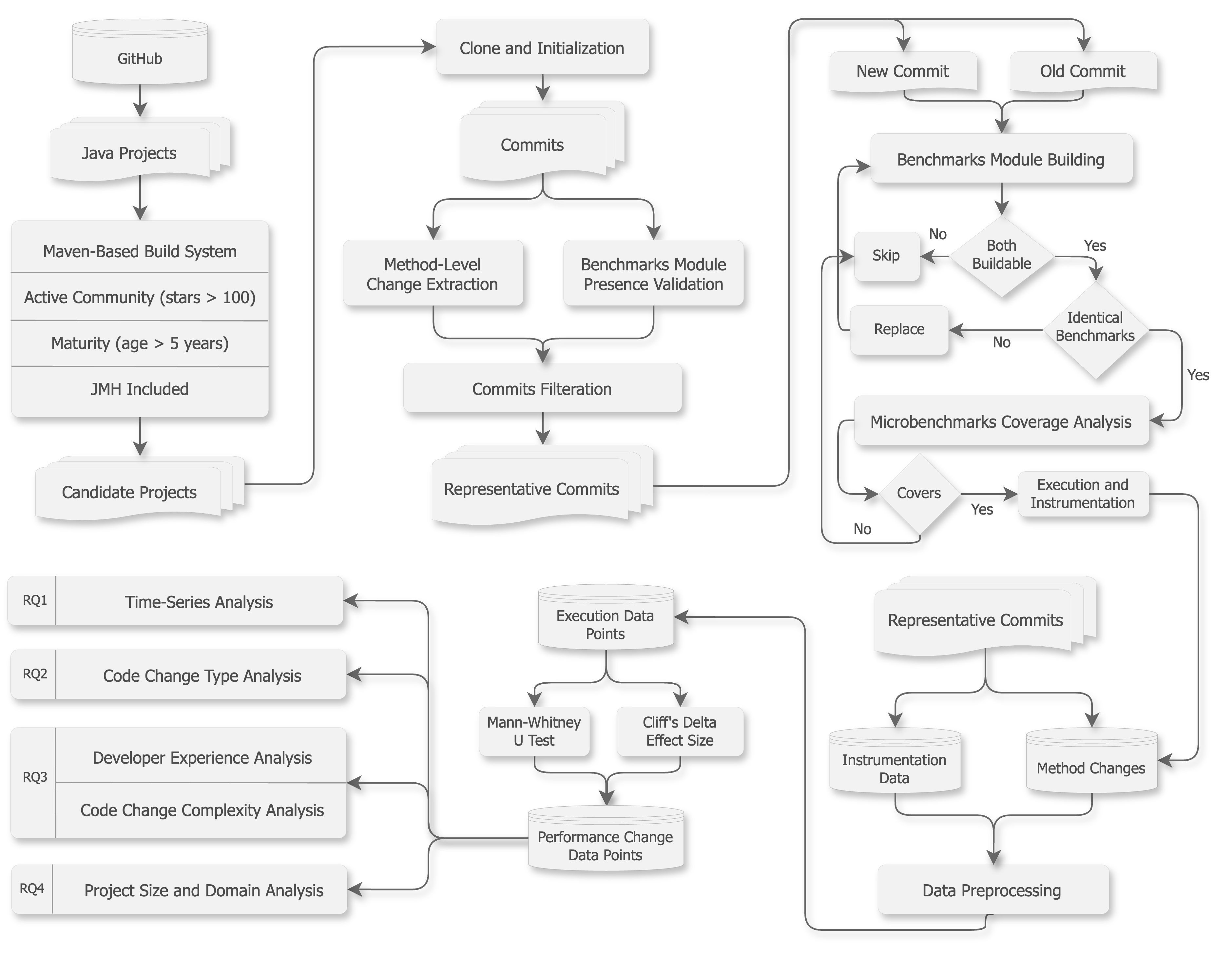}
    \caption{General overview of our study.}
    \label{fig:overview}
\end{figure*}

\subsection{Project Selection}

To create a diverse and representative dataset, we selected various Java repositories. We chose Java projects due to their abundant availability on GitHub, active developer community providing diverse code changes to analyze, and robust benchmarking tools such as JMH that enable precise performance measurements for our microbenchmarking study.  Each project's primary build system had to be based on Maven since it is the most common build system among well-developed Java repositories~\citep{kula2018developers, soto2019emergence, mitropoulos2014bug}, ensuring consistency in the build process. Additionally, similar to previous work~\citep{traini2021software}, each repository needed to have at least 100 stars, indicating community interest and active maintenance. We also combined this criterion with the maturity of the projects (i.e., developing for more than 5 years). Furthermore, the projects were required to include the Java Microbenchmark Harness (JMH) module in their repository, allowing us to use performance benchmarking techniques to assess the impact of method changes.

We chose to use microbenchmarks rather than unit tests for performance evaluation since they are tailored for measuring the performance of small code units with high precision, capturing detailed metrics such as execution time and resource usage~\citep{stefan2017unit, leitner2017exploratory, georges2007statistically}. Also, they are mostly designed to represent real-world usage scenarios~\citep{laaber2019software}. Meanwhile, unit tests are generally designed to verify functional correctness and are therefore generally not suitable for precise performance measurements due to potential overhead and lack of control over execution environments. Additionally, microbenchmarks offer greater statistical confidence through configurable parameters such as execution repetitions, sampling frequency, and fine-grained profiling capabilities, all of which contribute to more robust analysis of performance variations~\citep{georges2007statistically, alcocer2019performance, traini2021software}. By employing microbenchmarks, we ensure that our performance assessments of method changes are reliable and reflect true performance characteristics.

Applying these criteria, we obtained an initial set of Java projects through GitHub's API by querying for repositories that matched our predefined criteria. We sorted these based on the projects' popularity (measured by stars) and also cross-referenced our selection with projects analyzed in a well-studied previous work~\citep{traini2021software}, which focused on analyzing the performance change impact of software refactoring. This process yielded 23 Java projects. However, since these projects had to be aligned with our benchmarking pipeline (see~\ref{subsec:benchmarking}), this was filtered down to {\numProjects} final projects to ensure relevance while maintaining diversity in size, complexity, and domain. This selection enhances the generalizability of our findings across different applications. Furthermore, similar well-established studies in this field have analyzed a comparable number of projects (e.g., 9 to 19 projects)~\citep{alcocer2019performance,alcocer2015tracking,sandoval2016learning,laaber2020dynamically,chen2019analyzing,reichelt2019peass}, indicating that our selection aligns with standard research practices.

\subsection{Data Collection and Commit Filtering}

For each project, we iterated through its full Git commit history to investigate how code changes contribute to performance fluctuations over time. We processed each commit individually and applied the following criteria to identify the candidates for further steps. Specifically, through the entire set of commits, we ignored the merge commit from analyzing. 
Also, since benchmark modules are often added to Java projects after a specific point during their development
lifecycle~\citep{traini2021software}, we had to ignore those commits that were introduced before the presence of benchmarks.
Commits without code modifications to \texttt{.java} files were skipped since our analysis relies on benchmarking method changes that were done on Java codes. Finally, we excluded changes made exclusively to test files, focusing on the production codebase of the program.

After filtering, for each project, we had a set of commits that contained valid code modifications, specifically on methods, which we used for the code change extraction step.

\subsection{Method-Level Code Change Extraction}

To normalize and analyze code changes with precision, we implemented several specialized source code processing techniques that enabled accurate method tracking and comparison across versions.

We employed \textit{srcML}~\citep{collard2013srcml} to convert Java source code into XML format, specifically to facilitate method signature extraction and code cleanup. This process enabled us to remove all comments from the code and extract the names of all changed methods (retrieved from Git diffs), along with their identifiers, return types, access types, and full declaring class identifiers. These method signatures were essential for our subsequent instrumentation process, allowing us to target methods for performance measurement.

To perform deeper code analysis, we developed a custom Java code analysis tool called \textit{Java Parser Buddy} (JPB)~\footnote{\jpbURL}, our recently developed tool built on top of \textit{javaparser}~\citep{javaparser}, a mature, well-known Java code analysis library. Rather than reinventing the wheel, JPB serves as a bridge between our study and javaparser's sophisticated capabilities, enabling more straightforward code analysis. JPB constructs the Abstract Syntax Tree (AST) of the provided Java code and performs various analyses. In our case, we used JPB for two key purposes: 1) Identifying and replacing all string literals within method blocks with a constant value (i.e., "X"). This allowed for a more accurate comparison between two versions of a method, as many code changes only affected string literal values, which we chose to ignore due to their minimal impact on performance. 2) Tokenizing the code into meaningful elements such as keywords, operators, and identifiers. This helped us compare method changes more precisely to determine whether a substantive code modification had occurred.

To track methods that underwent structural changes while maintaining their functional purpose, we used \textit{RefactoringMiner}~\citep{Tsantalis:TSE:2020:RefactoringMiner2.0} to detect and classify refactorings at the method level. This analysis was crucial for establishing method identity across versions despite signature changes, ensuring we could properly attribute performance changes to the correct method evolution path. \textit{RefactoringMiner} identified method-level refactorings such as \textit{Rename Method}, \textit{Move Method}, \textit{Pull Up Method}, and \textit{Push Down Method}, each altering a method's signature and potentially impacting performance. Accordingly, we systematically mapped refactored methods before and after modification.

Combining Git diffs, \textit{srcML}, JPB, and \textit{RefactoringMiner}, we reassessed the set of considered commits and removed those that didn't include valid method-level code changes (i.e., those that were either out of methods' scope or didn't include proper code changes).

\subsection{Identifying Potentially Buildable Commits}
\label{subsec:identify-representative-commits}

We define a \ul{buildable commit} as one that can be successfully compiled and executed within its target environment. Our analysis focuses exclusively on \ul{potentially buildable commits} (i.e., those demonstrating empirical evidence of successful compilation in their historical context, such as verified through GitHub CI pipelines). This approach is motivated by the observation that commits with prior build failures exhibit a high probability of continued compilation issues, rendering them unsuitable for systematic analysis and resulting in inefficient allocation of computational resources.

Accordingly, after constructing the cleaned set of commits containing method-level changes, we applied further pruning to ensure data quality and optimize the resources and time required for building the projects. First, for projects using Continuous Integration (CI) pipelines on GitHub, we checked the build status of each commit using GitHub's API endpoints. If a commit had a build status that was not successful, we excluded it from the dataset. Additionally, we verified the presence of essential build configuration files, particularly the \texttt{pom.xml} file. Commits lacking this file were excluded, as they could not be properly built. Finally, we identified the required Java version for building the project at each commit. If the required version was not specified, we defaulted to Java 8.

This pruning process yielded a final refined set of candidate commits suitable for our benchmarking and performance analysis. Each commit in the final set included at least one method-level code change and could be \underline{potentially} built using the appropriate Java version. Table \ref{tab:projects-details} provides detailed information about each project, along with the corresponding statistics pertinent to our analysis.

\subsection{Sampling Representative Commits}

Due to computational constraints, analyzing all the candidate commits for each project was impractical, as each commit required multiple resource-intensive steps: verifying build success for both the commit and its predecessor, identifying which microbenchmarks covered the changed methods, and executing these benchmarks to collect extensive performance data. To address this, we employed a sampling strategy to select a representative subset of candidate commits for each project. We assigned higher sampling weights to commits containing more method changes, prioritizing them in our execution queue. This approach focused on commits likely to provide a larger set of method changes and more pronounced performance fluctuations, thus creating a more complete dataset. The rationale behind this approach lies in the limited coverage of JMH microbenchmarks, which can lead to scenarios where certain modified methods in each commit cannot be invoked by the microbenchmarks. Therefore, commits with a greater number of changed methods increase the likelihood of having more executable methods covered by the microbenchmarks, enhancing the effectiveness of the analysis.

In our case, using an appropriate sampling strategy was crucial, as traditional statistical methods like random sampling were not suitable for our case. While we initially filtered unbuildable commits based on GitHub CI workflow status (see~\ref{subsec:identify-representative-commits}), this did not guarantee actual buildability on our infrastructure due to various reasons, such as environment dependencies, network-specific configurations, and temporal third-party service availability. If a sampled commit turned out to be unbuildable (e.g., the \texttt{mvn install} command was failing), it would need to be replaced, rendering statistical random sampling ineffective. To overcome this, we employed a more tailored statistical and systematic sampling technique~\citep{cochran1977sampling}.
Our sampling strategy aimed to select candidates (sampled) from representative commits (from~\ref{subsec:identify-representative-commits}) while addressing constraints like buildability and benchmark coverage. We began by calculating the required sample size, $S$, for each project individually using standard statistical methods (95\% confidence level and 5\% margin of error) to ensure the representativeness of the selected samples. The systematic sampling interval, $k$, was then determined by dividing the total number of sorted commits $N$ by the required sample size $S$ for each project. Starting from a randomly selected position within the first interval of our sorted commit list, we attempted to build and execute the benchmarks at regular intervals of $k$ commits. This process continued until either the required number of samples (i.e., $S$) was obtained or all commits in the sorted list were analyzed. In cases where the required sample size could not be achieved, it was due to the failure to successfully analyze most of the commits (see~\ref{subsec:benchmarking} for more details).

\subsection{Benchmark Building and Execution}
\label{subsec:benchmarking}

For each sampled commit, we had to build the project's benchmarks and execute them to evaluate the performance fluctuations of method changes.

\subsubsection{Building the Benchmarks}

We manually reviewed the build configurations and steps for each project using their GitHub repository documentation. In most cases, the JMH benchmarks were defined as a separate module in \texttt{pom.xml}, requiring us to use the command \texttt{mvn -pl BENCH\_MODULE -am clean package} to build and package them correctly. When the project's configuration did not specify the benchmarks module separately, we first built the entire project using \texttt{mvn clean install} and then packaged the benchmarks module separately by running \texttt{mvn clean package} in the benchmark directory. In both scenarios, this process resulted in a proper executable \texttt{.jar} file that contained all the JMH microbenchmarks.

When the build process failed due to missing dependencies or configuration issues, we attempted to resolve these by adjusting build parameters or modifying environment settings, such as changing the Java version (switching between Java 8, 11, 17, and 21) or using the project's Maven wrapper (if available) instead of the system's global Maven. Also, for certain projects, we manually investigated their most common build issues, which necessitated implementing hard-coded adjustments in their build configurations to ensure successful execution. These modifications primarily addressed deprecated APIs or outdated dependencies that prevented proper building under our environment configurations. If the build failure persisted despite these efforts, we excluded the commit from further analysis.

\subsubsection{Dealing with Changed Microbenchmarks}

For commits where JMH benchmarks were modified with other code changes, we opted to use the older version of the benchmarks (i.e., from the previous commit) to ensure a fair comparison between code versions before and after the commit. This approach eliminates potential performance variations caused by benchmark modifications rather than actual code changes under test. We replaced the new benchmarks with the previous version and rebuilt them. If this procedure failed, we used the newer version of the benchmarks for both commits. If both replacement attempts were unsuccessful, we ignored the commit from the analysis, which could lead to inaccurate results.

\subsubsection{Selecting Microbenchmarks Covering Changed Code}

Since executing all microbenchmarks was not feasible for each single commit, we chose to run only those targeting the method changes in each commit. By listing all available microbenchmarks using \texttt{java -jar BENCHMARK.jar -l}, we executed each microbenchmark individually with minimal settings (1 fork, 1 iteration, 1 second per iteration, no warm-up). We also attached a lightweight instrumentation agent, \textit{Java Instrumentation Buddy} (JIB)~\footnote{\jibURL}, to the running JVM to collect coverage data for each microbenchmark, allowing us to identify those that specifically invoked the modified methods.

When candidate microbenchmarks overlapped with targeting changed methods (i.e., multiple microbenchmarks had common invoked methods), we applied an optimization algorithm to select a subset that maximized coverage while minimizing the total number of microbenchmarks executed (i.e., running the least number of microbenchmarks while covering all of the possible method changes). Additionally, we stored the coverage information for each benchmark version, allowing us to reuse this data later without the need to re-run the benchmarks if their definitions remained unchanged. All of the details and codes are in our replication package.

If no benchmarks targeted the changed methods, we excluded the commit from analysis, as we could not execute the changed methods and evaluate their performance. In fact, this phenomenon has been investigated in other works, and it is determined that performance benchmarks typically have lower coverage rather than unit tests~\citep{traini2021software, stefan2017unit, laaber2018evaluation, chen2017exploratory}.

\subsubsection{Executing Benchmarks with Instrumentation}

Instrumentation is a common practice to obtain the execution data of methods within programs~\citep{gebai2018survey, desnoyers2006lttng}.
With a map of microbenchmarks and their target methods, we executed them for both the current commit and its predecessor to assess performance fluctuations. We configured the execution with 3 forks and 5 iterations (each lasting 10 seconds) to balance execution time and statistical reliability (i.e., 15 different iterations in total), following established practices for Java performance evaluation~\citep{georges2007statistically, costa2021whats, laaber2022effective}. To instrument the modified methods at runtime, we utilized a custom-developed tool called Java Instrumentation Buddy (JIB), built using the widely recognized Java bytecode modification library, \textit{Byte-Buddy}~\citep{bytebuddy}. JIB was specifically designed to minimize overhead (generally less than 2\%) while accurately instrumenting the modified methods and recording the resulting data in trace files for subsequent analysis. By executing benchmarks under identical conditions for both commits, we ensured that any observed performance differences were attributable to code changes rather than environmental factors.

\subsection{Performance Data Analysis}

We performed a thorough examination of the instrumentation data to evaluate the impact of method-level code changes on execution times. Our instrumentation collected detailed traces from each modified method, capturing metrics such as start and end times for each method. This raw data formed the foundation for constructing execution time distributions and performing subsequent statistical analyses.

In the first step, we extracted the execution times for each modified method into distinct distributions. This allowed us to visually and statistically inspect the variability and central tendencies in method performance. To ensure that external or transient factors (e.g., system interrupts, background processes) did not distort our findings, we applied the Interquartile Range (IQR) method to remove outliers from each distribution. Specifically, we calculated the first (Q1) and third (Q3) quartiles and excluded all data points that lay outside the range \([Q1 - 1.5 \times IQR,\; Q3 + 1.5 \times IQR]\). By eliminating these extreme values, we aimed to capture a more accurate representation of the typical execution times and prevent sporadic fluctuations from skewing the analysis.

\textbf{Performance Change Detection:} To systematically assess whether there were meaningful performance differences between the old and new versions of each method, we conducted the Mann-Whitney U-test~\citep{nachar2008mann} on the cleaned execution time distributions. The Mann-Whitney U-test is a nonparametric statistical test that does not assume normality in the underlying data, making it especially suitable when the distribution of execution times may be skewed or contain multiple modes. Whenever the Mann-Whitney U-test detected a statistically significant difference (p $<$ 0.05), we calculated Cliff's delta~\citep{cliff2014ordinal} to determine the effect size. Cliff's delta quantifies the degree of overlap between two distributions, offering additional insight into whether the observed performance shift is practically meaningful. We established the standard threshold of $|\text{effect size}| \geq 0.147$ to distinguish between negligible and meaningful performance changes, ensuring our analysis focuses on practically significant impacts rather than statistical noise.

\textbf{Categorical Analysis:} Chi-square tests of independence~\citep{snedecor1989statisticalchi} were employed to assess whether the distribution of performance outcomes (improvement, regression, unchanged) differs significantly across categorical variables such as project lifecycle stages, code change types, developer experience levels, project domains, and size categories. Fisher's exact tests~\citep{fisher1970statistical} were applied for pairwise comparisons between specific categories when contingency table cell counts were small, providing more accurate p-values under these conditions.

\textbf{Group Comparisons:} One-way ANOVA~\citep{snedecor1989statisticalchi} was used to test for significant differences in continuous variables (such as absolute effect sizes) across multiple groups, particularly when comparing developer experience categories and project characteristics. Tukey's HSD post-hoc tests~\citep{tukey1949comparing} were applied following significant ANOVA results to identify which specific group pairs differ significantly while controlling for multiple comparisons. For non-normally distributed data, Kruskal-Wallis tests~\citep{kruskal1952use} served as the nonparametric alternative to ANOVA, particularly when comparing effect size magnitudes across code change complexity categories and project domains.

\textbf{Correlation and Interaction Analysis:} Pearson correlation coefficients~\citep{cohen2013applied} were calculated to assess linear relationships between continuous variables, such as code change complexity scores and performance impact magnitudes. Spearman rank correlations~\citep{spearman1961proof} were used to detect monotonic relationships that may not be strictly linear. Two-way ANOVA~\citep{montgomery2017designanovatwo} was employed to detect interaction effects between categorical factors (e.g., project domain and size), testing whether the effect of one factor varies depending on the level of another factor.

\textbf{Reliability and Agreement Measures:} Cohen's Kappa~\citep{cohen1960coefficient} was calculated to quantify inter-rater agreement during the code change classification process, ensuring that categorical assignments were reliable and consistent across multiple evaluators. Values of $k \ge 0.8$ were considered to indicate excellent agreement according to established thresholds~\citep{mchugh2012interrater}.

\textbf{Effect Size Quantification:} Beyond statistical significance, we systematically measured practical significance through multiple effect size measures. Cohen's d~\citep{cohen2013statistical} was calculated for comparing means between two groups, with thresholds of 0.2 (small), 0.5 (medium), and 0.8 (large) effect sizes. Cramér's V~\citep{cramer1999mathematical} was computed to measure the strength of association between categorical variables, providing insight into practical significance beyond statistical significance. For all effect size measures, we reported confidence intervals at the 95\% level and applied appropriate corrections for multiple comparisons when conducting simultaneous tests across multiple categories.

This comprehensive statistical framework ensures that our findings represent both statistically significant and practically meaningful differences in performance patterns across the various factors examined in our research questions.

\subsection{Dataset Compilation}

Finally, we have provided a complete dataset of performance changes of {\numMethodChanges} method changes in {\numBenchmarkedCommits} commits that were derived from our {\numProjects} projects. The dataset may be found in our replication package

%% file: sections/evaluations.tex
\subfile{../tables/projects_details}

\subsection{Experiment Setup}
We developed a pipeline that automates the build process and conducts performance benchmarking for Maven-based Java projects, ensuring consistent and efficient evaluation of application performance. We deployed it across multiple virtual machines simultaneously, each running an individual project. We used Google Cloud Platform Compute Engine to provision our virtual machines, each equipped with an 8-core CPU, 16\,GB of memory, and 250\,GB of SSD storage. To minimize performance variations due to external factors, we did not interact with the machines during the benchmarking process. For each measurement (i.e., running current and previous commit), we used 3 forks and 5 iterations (each taking 10 seconds) to account for measurement errors (i.e., 15 iterations in total)~\footnote{It is important to note that the number of iterations does not represent the total number of times a method has been executed across a commit and its predecessor. In fact, during each iteration, a method might be executed as few as once (e.g., the main method) or up to millions of times.}. The entire data collection process took approximately 80 machine days.

\subsection{RQ1: What are the patterns of performance changes in Java projects over time?}

\noindent\textbf{Motivation}

Understanding temporal patterns and magnitudes of performance changes is crucial for implementing effective continuous performance monitoring in software development. By analyzing when and how frequently performance regressions and improvements occur across project lifecycles, we can inform the design of automated performance testing strategies and identify critical periods requiring heightened attention. This analysis provides the empirical foundation for integrating performance validation into continuous integration pipelines and guides resource allocation for performance management activities.

\smallbreak

\noindent\textbf{Approach}

\ul{Performance Change Definition and Statistical Framework:} We formally define performance changes as follows: A performance improvement or regression occurs when the Mann-Whitney U-test indicates statistical significance ($p < 0.05$) and the absolute effect size (Cliff's Delta) is non-negligible ($|\text{effect size}| \geq 0.147$). Changes not meeting these criteria are classified as ``unchanged performance.'' This definition ensures that our analysis focuses on practically meaningful performance variations rather than statistical noise. We employ the Mann-Whitney U-test due to its non-parametric nature, making it suitable for performance data that may not follow normal distributions, while Cliff's Delta provides robust effect size quantification independent of sample size variations.

\ul{Project Timeline Normalization and Phase Division:} To analyze evolution patterns across projects with varying lifespans and development velocities, we normalized each project's timeline from 0 (project start) to 1 (project end or current time) using min-max normalization technique, following approaches used in longitudinal software evolution studies~\citep{mockus2000empirical, godfrey2000evolution, lehman1980programs}. This enables fair comparison across projects with different development spans, from months to decades. We divided this normalized timeline into three equal phases: \textit{Early Stage} (0-0.33), \textit{Middle Stage} (0.33-0.67), and \textit{Late Stage} (0.67-1.0), allowing systematic analysis of how performance change patterns evolve throughout project maturity cycles. While project development phases may not align perfectly with equal temporal divisions, this approach provides a consistent framework for comparative analysis across diverse projects~\citep{hassan2008road, bird2009fair}.

\ul{Temporal Pattern Analysis and Volatility Quantification:} Rather than using moving averages of effect sizes, which is statistically inappropriate for effect size data, we calculated the proportion of commits resulting in performance improvements, regressions, and unchanged performance across normalized time periods. To quantify the consistency of performance change patterns over time, we calculate a volatility metric that measures the variance in performance instability rates across sliding time windows throughout each project's development lifecycle. Specifically, we divide each project's timeline into overlapping 20\% windows (stepped by 10\% increments) and calculate the instability rate (percentage of changes causing performance impacts) for each window. Volatility is computed as the statistical variance of these instability rates across all time windows, where higher values indicate erratic performance patterns and lower values suggest consistent behavior throughout development.

\begin{figure}
    \centering
    \includegraphics[width=0.95\columnwidth]{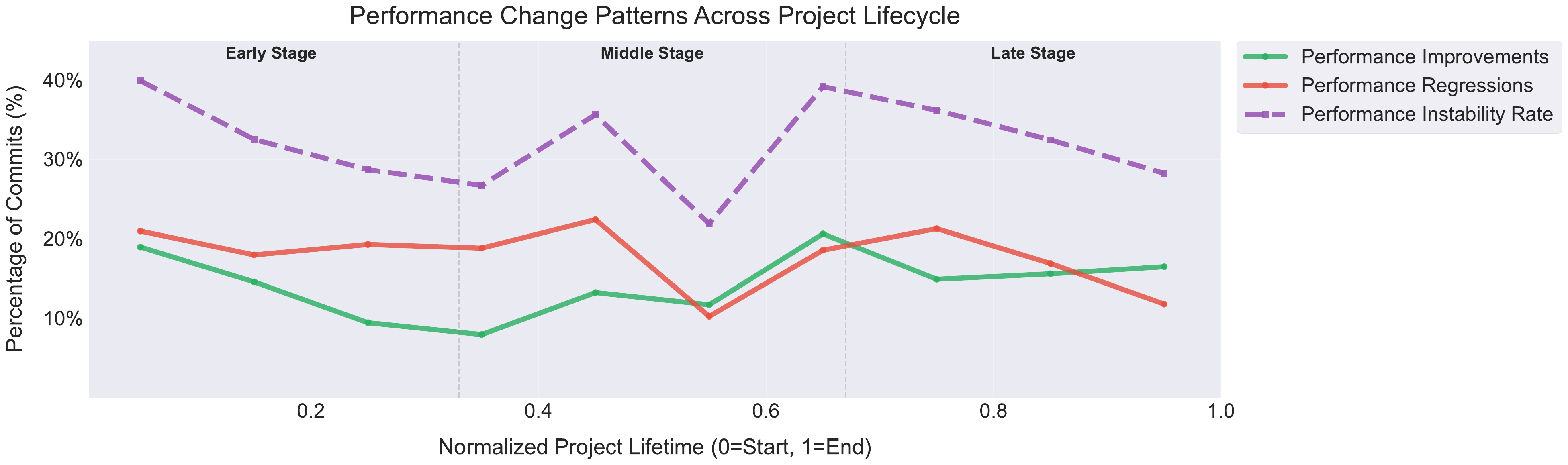}
    \captionsetup{justification=centering}
    \caption{Performance change proportions across normalized projects' lifecycle. The instability rate represents the percentage of commits causing any performance change (improvements + regressions).}
    \label{fig:rq1_effect_size_evolution}

    \smallbreak

    \centering
    \includegraphics[width=0.95\columnwidth]{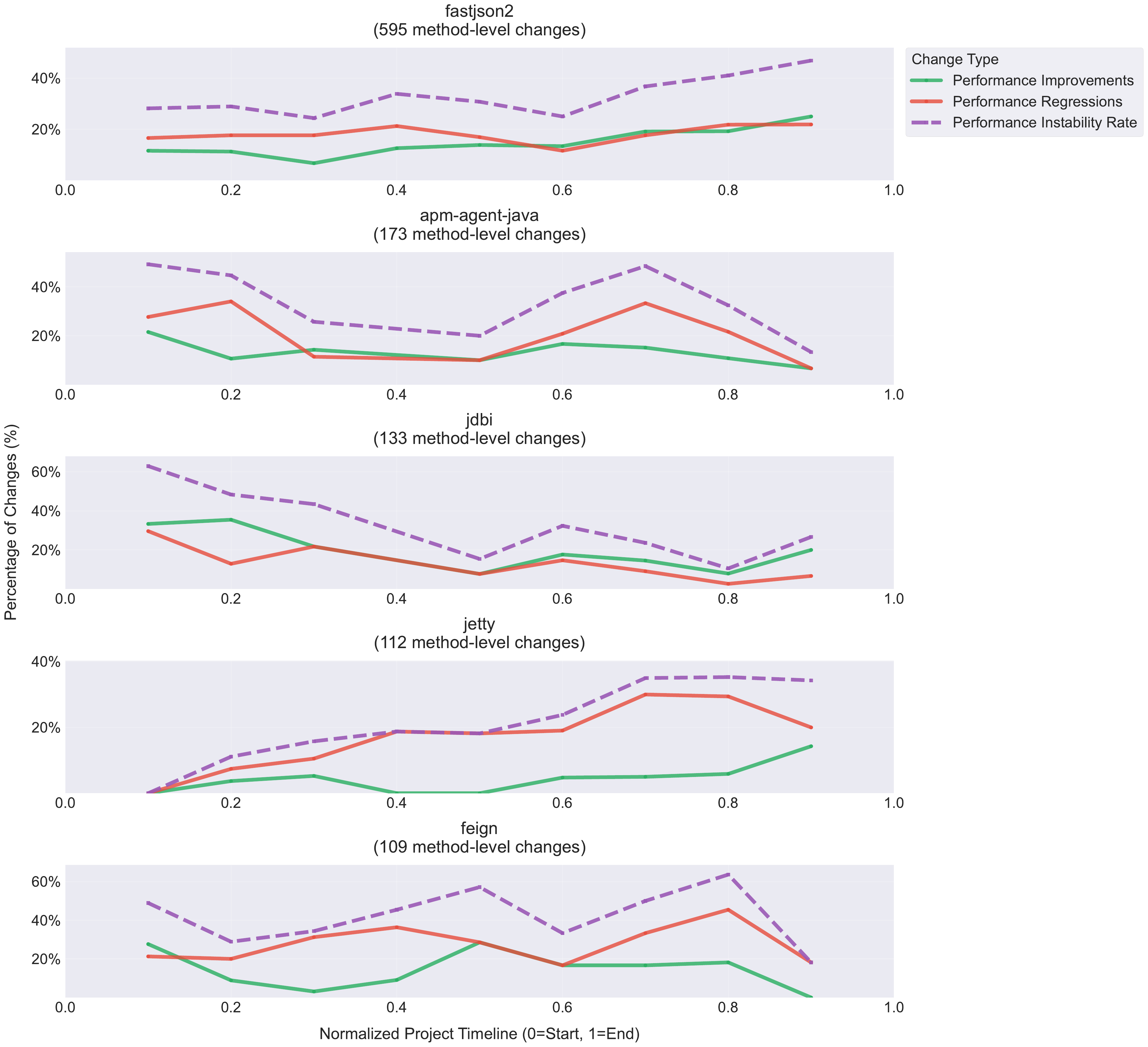}
    \captionsetup{justification=centering}
    \caption{Performance change patterns across individual project lifecycles for the five most data-rich projects. Each panel shows the proportion of performance improvements, regressions, and overall instability rate over the normalized project timeline.}
    \label{fig:rq1_effect_size_evolution_individual}
\end{figure}

\smallbreak

\noindent\textbf{Results}

\ul{Performance regressions and improvements are prevalent throughout the software lifecycle, with significant implications for continuous integration practices.} Our analysis of {\numMethodChanges} method-level changes across {\numBenchmarkedCommits} commits reveals that performance changes occur consistently throughout project development. The results show that 212 changes resulted in performance improvements (14.2\% of total), 275 changes led to performance regressions (18.5\% of total), and 1,002 changes exhibited unchanged performance (67.3\% of total). Table~\ref{tab:rq1_performance_distribution} illustrates the distribution of method-level code changes in this study.

\subfile{../tables/rq1/overall_distribution}

This distribution demonstrates that nearly one-third (32.7\%) of method-level code changes result in measurable performance impacts, with regressions being 1.3 times more frequent than improvements. The prevalence of performance-affecting changes throughout the development lifecycle strongly supports the integration of automated performance testing in continuous integration pipelines, as developers cannot rely on functional tests alone to detect performance regressions. Figure~\ref{fig:rq1_effect_size_evolution} illustrates the temporal distribution of these performance changes, showing consistent occurrence of both improvements and regressions across the analyzed timeframe.

\ul{Individual projects exhibit substantial variation in performance change patterns, demonstrating the need for tailored monitoring strategies.}
Analysis of the five projects with the most method-level changes reveals significant heterogeneity in performance stability characteristics (Figure~\ref{fig:rq1_effect_size_evolution_individual}). Performance instability rates vary considerably across projects, ranging from 23.2\% (\textit{jetty}) to 42.2\% (\textit{feign}), with a standard deviation of 6.4\% indicating meaningful variation. This 1.8-fold difference in instability rates demonstrates that project-specific factors substantially influence performance change frequency.

Beyond overall instability rates, projects exhibit distinct volatility patterns in their performance evolution. High-volatility projects such as \textit{jdbi} (volatility = 273.5) and \textit{feign} (volatility = 188.2) demonstrate erratic performance change patterns with peak instability rate reaching up to 63.6\%, while low-volatility projects like \textit{fastjson2} (volatility = 50.6) and \textit{jetty} (volatility = 129.2) maintain more consistent performance change rates throughout their development cycles. This variation suggests that some projects require more intensive performance monitoring due to their inherently unstable performance characteristics.

Projects also demonstrate contrasting lifecycle patterns that further support the need for differentiated monitoring approaches. Early-heavy projects (\textit{apm-agent-java}, \textit{jdbi}, \textit{feign}) exhibit higher performance instability during initial development phases, with early-stage instability rates of 40.8\%, 56.2\%, and 43.0\%, respectively, followed by stabilization in later stages. Conversely, late-heavy projects (\textit{fastjson2}, \textit{jetty}) show increasing performance activity toward project maturity, with late-stage instability rates of 41.0\% and 39.5\% respectively. These opposing patterns indicate that performance management strategies should be adapted based on both project maturity and historical performance behavior, rather than applying uniform monitoring approaches across all projects.

\begin{figure}
    \centering
    \includegraphics[width=\columnwidth]{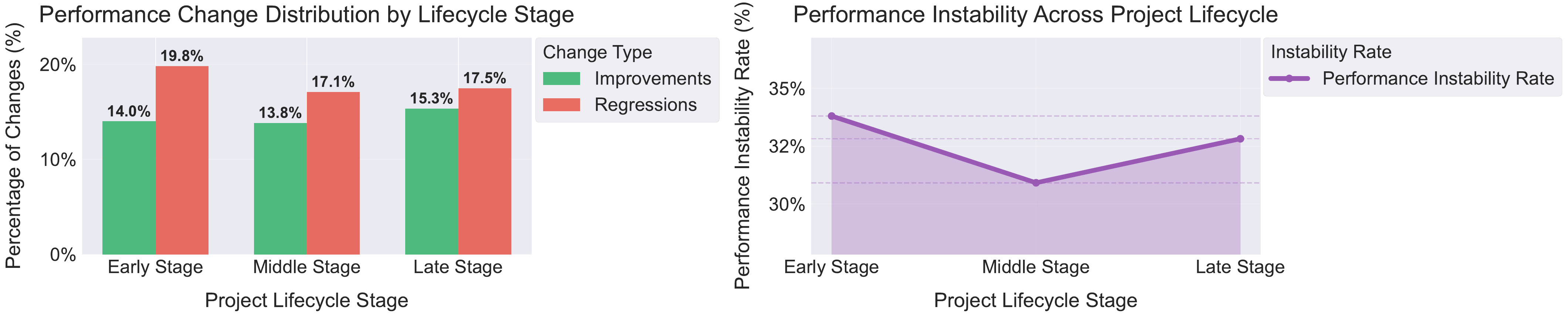}
    \captionsetup{justification=centering}
    \caption{Performance change patterns across project lifecycle stages. Left panel shows the distribution of performance improvements and regressions for each development stage (Early: n=707, Middle: n=456, Late: n=326). Right panel displays the performance instability rate (improvements + regressions) across lifecycle stages, revealing a modest U-shaped pattern with early-stage instability at 33.8\%, middle-stage at 30.9\%, and late-stage at 32.8\%.}
    \label{fig:rq1_lifecycle_comparison}
\end{figure}

\ul{Performance instability exhibits modest but consistent patterns across project lifecycle stages.}
Figure~\ref{fig:rq1_lifecycle_comparison} reveals that performance instability rates follow a subtle U-shaped progression: early-stage development (33.8\% instability, n=707), middle-stage development (30.9\% instability, n=456), and late-stage development (32.8\% instability, n=326). Statistical analysis confirms these differences are not significant ($\chi^2$ = 2.015, p = 0.73, Cramér's V = 0.026), indicating consistent monitoring requirements throughout development. However, the improvement-to-regression ratio improves with maturity (0.71 → 0.81 → 0.88), suggesting better performance outcomes in later stages.

\begin{figure}
    \centering
    \includegraphics[width=\columnwidth]{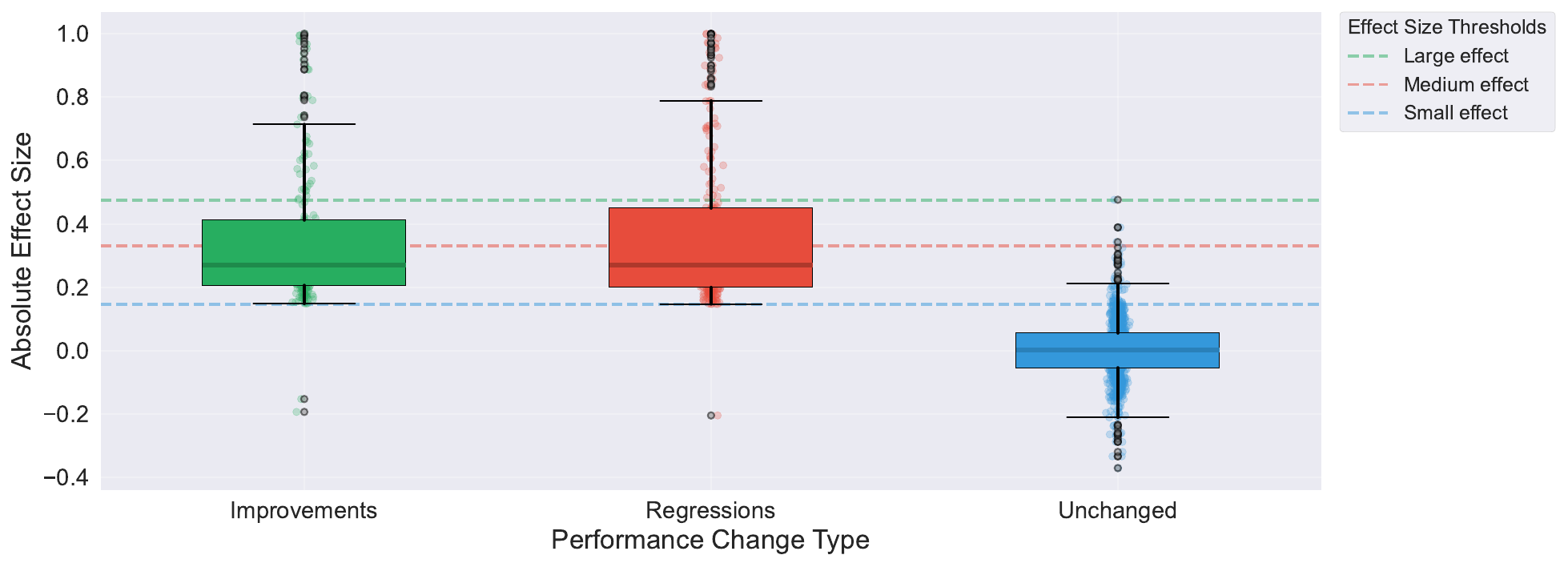}
    \captionsetup{justification=centering}
    \caption{The performance change effect size categories distribution.}
    \label{fig:rq1_distribution}
\end{figure}

\ul{Performance improvements and regressions exhibit similar effect size distributions with no statistically significant difference in magnitude.} Statistical comparison reveals balanced performance change patterns between improvements and regressions:

\begin{itemize}
\item \textbf{Improvement effect size}: Mean = 0.395 ($\pm$0.270 $std$), Median = 0.289
\item \textbf{Regression effect size}: Mean = 0.429 ($\pm$0.288 $std$), Median = 0.296  
\item \textbf{Mann-Whitney U-test}: $p$ = 0.508, indicating no statistically significant difference
\item \textbf{Cohen's d}: -0.120 (negligible effect size difference)
\end{itemize}

As shown in Figure~\ref{fig:rq1_distribution}, both improvements and regressions demonstrate similar central tendencies and variability, indicating that the magnitude of performance impacts is comparable regardless of the direction. While regressions show a slightly higher mean effect size, this difference is not statistically significant, suggesting that both positive and negative performance changes should be monitored with equal attention and similar thresholds.

\begin{figure}[]
    \centering
    \includegraphics[width=\columnwidth]{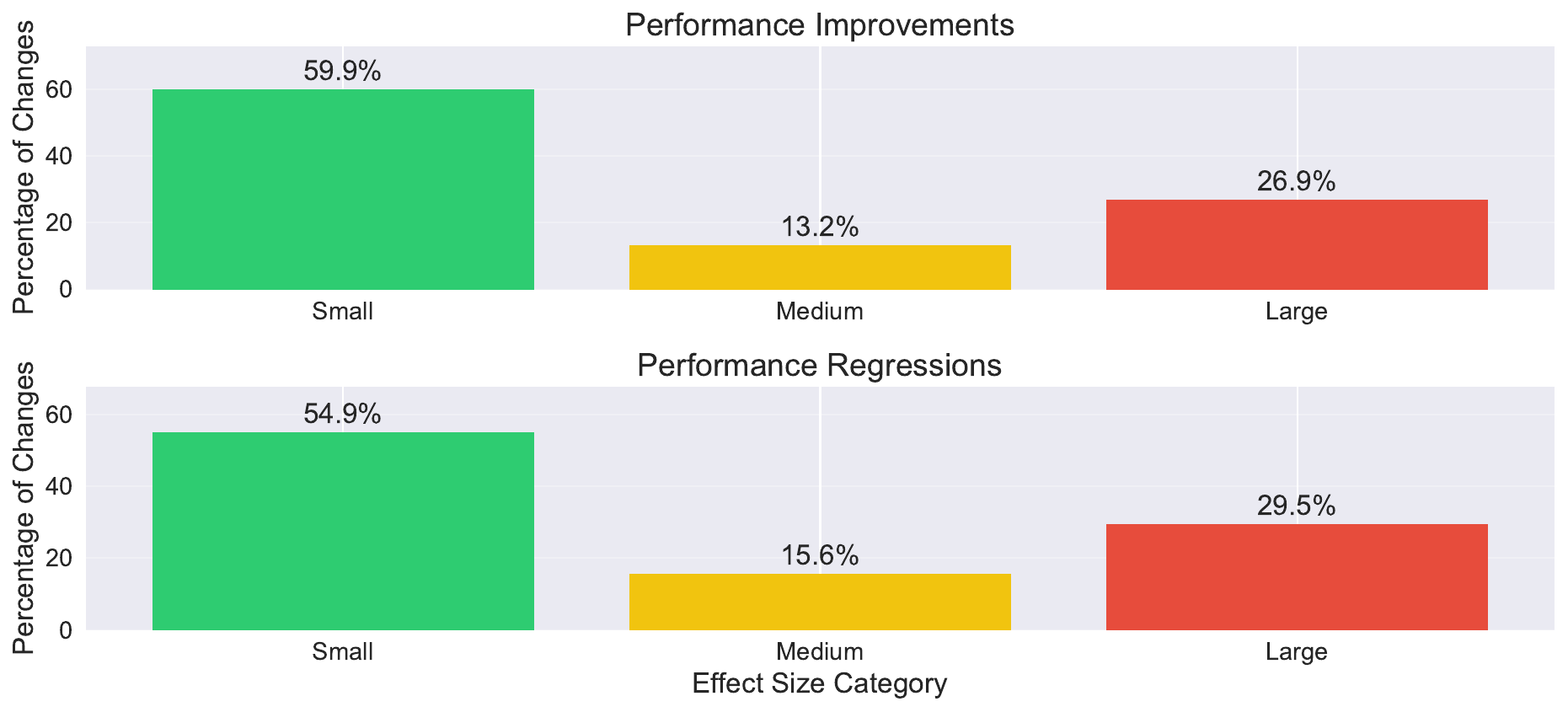}
    \captionsetup{justification=centering}
    \caption{The distribution of performance change effect size categories based on the performance change type.}
    \label{fig:rq1_categories}
\end{figure}

\ul{Most significant performance changes exhibit small effect sizes, with notable proportions reaching substantial magnitudes.} Analysis of effect size categories, illustrated in Figures~\ref{fig:rq1_distribution} and~\ref{fig:rq1_categories}, reveals that improvements follow this distribution: small effect sizes (0.147-0.33) account for 59.0\% of improvements, medium effect sizes (0.33-0.474) represent 13.2\% of improvements, and large effect sizes ($\geq$0.474) constitute 26.9\% of improvements. Similarly, regressions display a comparable pattern with small effect sizes comprising 54.9\% of regressions, medium effect sizes accounting for 15.6\% of regressions, and large effect sizes representing 29.5\% of regressions. This distribution highlights that while the majority of performance changes are modest in magnitude, a substantial portion (i.e., 29.5\% of regressions and 26.9\% of improvements) exhibit large effects. The similar distribution patterns between improvements and regressions suggest that both types of performance changes follow comparable magnitude patterns, supporting the need for balanced monitoring strategies, which is crucial for comprehensive performance management, as it enables teams to identify beneficial optimizations that can compensate for performance regressions in other code areas, maintain an optimization knowledge base for future development decisions, and detect when apparent improvements might actually indicate measurement errors or unintended side effects. Additionally, understanding improvement patterns helps teams validate the effectiveness of their optimization strategies and guides future performance enhancement efforts. The prevalence of small effect sizes (approximately 57\% of all significant changes) indicates that incremental performance optimizations are more common than dramatic improvements, reflecting iterative refinement processes in software development.

\begin{boxK}
    \textbf{RQ1 Conclusion and Implications} \\
    \ul{For Continuous Integration Implementation}: The finding that 32.7\% of method changes impact performance provides strong empirical justification for integrating automated performance testing into CI pipelines. Organizations should implement performance regression detection with appropriate thresholds based on effect size significance rather than arbitrary percentage changes.
    
    \ul{For Development Resource Allocation}: While early-stage development shows slightly higher regression rates (19.8\% vs 17.5\% in late stage), the modest variation across lifecycle stages suggests that performance engineering resources should be maintained consistently throughout project development rather than being concentrated in specific phases.
    
    \ul{For Risk Management}: The similar effect size distributions between improvements and regressions indicate that both positive and negative performance changes require equal monitoring attention. Teams should implement balanced performance review processes that detect both potential optimizations and performance degradations, as improvements can mask regressions in other code sections, help identify effective optimization patterns for replication, and ensure that performance testing captures the full spectrum of code change impacts rather than focusing solely on regression prevention.
    
    \ul{For Project-Specific Strategies}: The substantial variation in performance change patterns across individual projects emphasizes the need for tailored performance management approaches that account for project maturity, complexity, and historical behavior patterns.
\end{boxK}

\subsection{RQ2: How do different types of code changes correlate with performance impacts?}

\noindent\textbf{Motivation}

The relationship between specific method-level code modification types and their performance consequences remains poorly understood, despite its critical importance for development practices. While developers often hold intuitive beliefs about which changes are "risky" versus "safe" for performance, such as avoiding I/O modifications or prioritizing algorithmic optimizations, these assumptions lack empirical validation. Understanding how different categories of method-level code changes actually correlate with performance improvements and regressions can provide insights for enhancing development strategies, code review priorities, and performance testing approaches. By systematically analyzing the performance impact patterns across modification types, we can provide evidence-based guidance for performance-conscious development and challenge potentially unfounded conventional wisdom about code change risks.

\smallbreak

\noindent\textbf{Approach}

To understand how different types of method-level code changes correlate with performance impacts, we need a systematic approach to classify method-level modifications and analyze their performance consequences. While previous studies have established basic taxonomies of code changes~\citep{zhao2022large,sanchez2020tandem}, applying these classifications to performance analysis presents unique challenges. Existing taxonomies may not capture all performance-relevant modification patterns, and the relationship between change types and performance outcomes requires careful statistical analysis to avoid spurious correlations. Furthermore, the subjective nature of change classification demands rigorous inter-rater reliability validation to ensure meaningful results.

\subfile{../tables/rq2/code_change_categories}

\ul{Change Type Classification:} We developed a systematic change classification approach that balances comprehensiveness with reliability. Our methodology employed a hybrid labeling strategy that combined established taxonomies from the literature with empirical refinement based on our dataset characteristics. We began with predefined categories derived from previous studies~\citep{zhao2022large,sanchez2020tandem} but explicitly allowed for the addition of new categories when existing classifications proved insufficient to capture meaningful distinctions observed in our Java projects. This hybrid approach ensured comprehensive coverage of modification types while maintaining systematic rigor and avoiding forced categorization of genuinely novel patterns that could compromise analytical validity.

Our taxonomy builds upon established software evolution literature while introducing refinements for performance analysis. Core categories such as Algorithmic Changes, Control Flow modifications, and Data Structure changes follow foundational taxonomies from software maintenance research~\citep{barry1999detailed,zhao2022large}, while categories like Concurrency and API modifications align with recent performance-focused studies~\citep{zhao2022large,sanchez2020tandem}. We refined the Exception \& Return Handling category specifically for this study, as existing taxonomies typically subsume these changes under broader corrective maintenance classifications without recognizing their distinct performance implications.

For method changes exhibiting characteristics of multiple categories, we employed a multi-labeling strategy where all applicable labels were assigned, ensuring comprehensive representation of complex modifications while preserving the ability to analyze individual category impacts through data expansion techniques. This approach recognizes that real-world code changes often span multiple dimensions simultaneously and prevents the loss of analytical granularity that would result from forcing changes into single categories. Table~\ref{tab:code-change-categories} presents the taxonomy of code change types identified through our hybrid approach, capturing the diverse range of changes with potential performance implications.

\ul{Balanced Dataset Construction:} We constructed a balanced dataset for statistical analysis to prevent bias toward performance-altering changes that could skew our findings. We first identified {\numSignificantMethodChanges} significant performance changes (those meeting our Mann-Whitney U-test significance and effect size thresholds from RQ1), then randomly selected an equal number of non-significant changes as controls. This approach yielded a final dataset of {\numLabeledMethodChanges} labeled method changes, providing robust statistical power for comparative analysis while ensuring that our results reflect genuine performance impact patterns rather than artifacts of unbalanced sampling.

\ul{Three-Phase Labeling Process:} To ensure classification reliability and consistency, we implemented a structured three-phase labeling protocol on our balanced dataset. In the \textbf{calibration phase}, two authors independently labeled an initial sample of 275 randomly selected method changes, with sample size determined using a 95\% confidence level and 5\% margin of error to ensure adequate statistical power for agreement assessment. During the \textbf{consensus phase}, multiple rounds of discussion were conducted to resolve disagreements and establish consistent labeling criteria. New categories were added during these discussions when justified by recurring patterns not adequately captured by existing classifications, ensuring our taxonomy reflected the actual diversity of modifications in real software projects. In the \textbf{independent labeling phase}, each author proceeded to label the remaining method changes using the finalized taxonomy established during consensus building.

\ul{Reliability Validation:} To quantify the reliability of our classification process, we calculated Cohen's Kappa~\citep{cohen1960coefficient} on the entire independently labeled dataset, excluding only the initial calibration sample used for training and consensus building. The resulting $k = 0.96$ indicates excellent inter-rater agreement according to established thresholds~\citep{mchugh2012interrater}, confirming the consistency and reliability of our classification approach.

\begin{figure}[]
    \centering
    \includegraphics[width=\columnwidth]{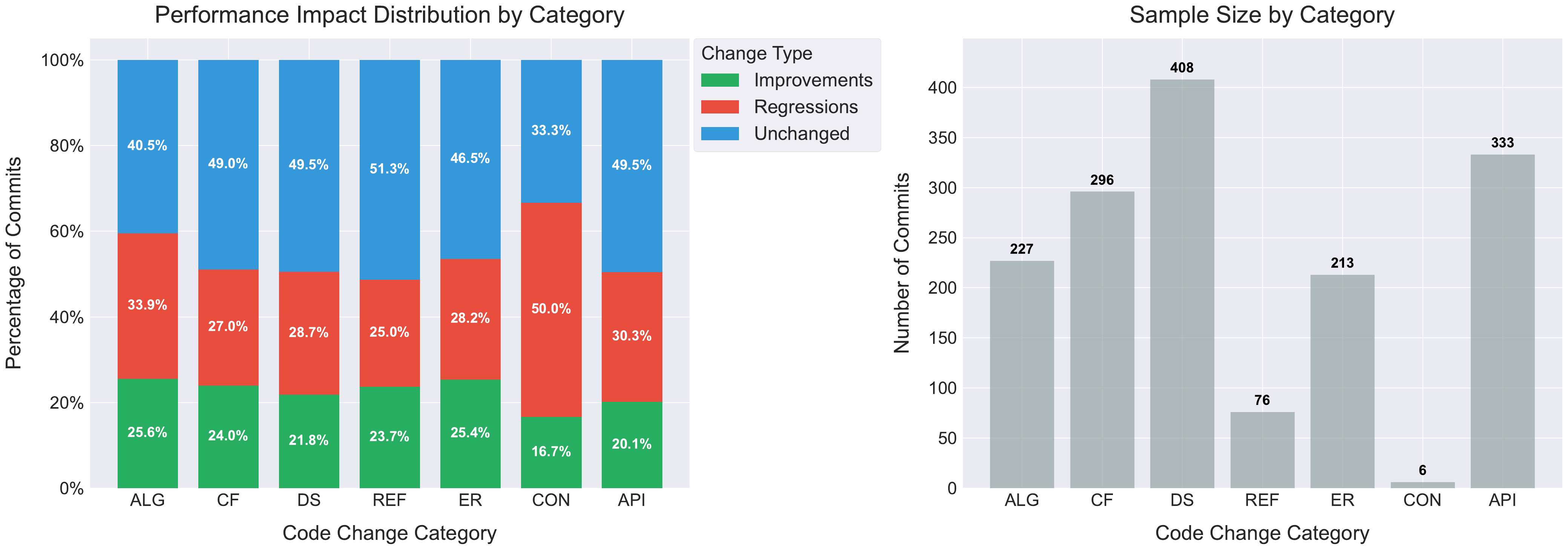}
    \captionsetup{justification=centering}
    \caption{Performance impact distribution and sample sizes by code change category. Left panel shows the proportion of commits resulting in improvements, regressions, and unchanged performance for each category. Right panel displays sample sizes, with annotations showing total commits analyzed per category.}
    \label{fig:rq2_proportional_impact}
\end{figure}

\smallbreak
\noindent\textbf{Results}

\ul{Performance impact patterns are statistically similar across all code change categories, challenging conventional assumptions about modification risk hierarchies}~\citep{mockus2000predicting,huang2014performance,kamei2013large}.
Statistical analysis of {\numLabeledMethodChanges} labeled method changes reveals no significant differences in performance impact distributions across code change categories ($\chi^2$ = 10.255, p = 0.594, Cramér's V = 0.057). This finding directly contradicts widespread developer intuitions~\citep{balijepally2015task,fritz2007does} about which types of code modifications are inherently more likely to cause performance regressions. As shown in Figure~\ref{fig:rq2_proportional_impact} and Table~\ref{tab:code-change-proportion}, all categories demonstrate remarkably similar patterns of improvements, regressions, and unchanged performance. The negligible effect size (Cramér's V = 0.057) indicates that code change category membership provides minimal predictive value for performance outcomes, suggesting that implementation quality, developer experience, and contextual factors may be far more influential than the fundamental type of modification being made. Analysis of effect size magnitudes shows remarkable consistency across code change categories, with 50-60\% of significant changes exhibiting small effects (0.147-0.33), 10-17\% showing medium effects (0.33-0.474), and 20-40\% demonstrating large effects ($\ge$0.474), indicating that when performance impacts occur, they tend to be practically significant rather than marginal, regardless of the underlying change type.

\subfile{../tables/rq2/code_change_type_proportions}

\ul{Algorithmic Changes are among the most likely to introduce significant performance changes, exhibiting the highest improvement potential while simultaneously carrying substantial regression risk, representing a high-stakes optimization opportunity.}
Algorithmic Changes demonstrate the highest improvement rate among all categories at 25.6\% (58/227 method changes), making them the most promising modification type for performance optimization efforts. However, they simultaneously exhibit the second-highest regression rate at 33.9\% (77/227 method changes), resulting in a risk ratio of 1.33 where regressions outnumber improvements by a meaningful margin. Figure~\ref{fig:cc-algorithmic-improvement} illustrates how algorithmic optimization can yield dramatic performance gains, such as replacing lazy request supplier patterns with direct buffer allocation, achieving an 85\% performance improvement. This dual nature characterizes algorithmic modifications as high-impact changes that can significantly enhance or degrade performance depending on implementation quality. The mean effect size of 0.470 for this category reinforces that algorithmic changes tend to produce substantial performance impacts in either direction, requiring careful implementation and comprehensive testing protocols.

\subfile{../figures/code-changes/algorithm_improvement}

\ul{Exception and Return Handling changes emerge as a surprisingly effective optimization category, outperforming more commonly targeted modification types.}
Exception and Return Handling modifications achieve the second-highest improvement rate at 25.4\% (54/213 method changes) while maintaining one of the most favorable risk ratios at 1.11, significantly outperforming categories that developers typically prioritize for performance optimization such as API/Library Call changes (20.1\% improvement rate). The 28.2\% regression rate (60/213 method changes) is below the sample average, indicating that modifications to exception handling patterns and return logic represent a consistently successful optimization approach. However, as demonstrated in Figure~\ref{fig:cc-exception-regression}, adding comprehensive exception handling and debugging capabilities can introduce performance overhead, with SSL debugging enhancements causing a 3.1\% regression due to the computational cost of comprehensive error handling and logging operations. This finding suggests that developers may be overlooking a highly effective optimization category by focusing primarily on algorithmic changes and API optimizations while neglecting the performance potential of exception handling improvements and return path optimizations. The mean effect size of 0.476 for this category, which is the highest among all types, indicates that successful optimizations in exception and return handling tend to produce meaningful performance gains, suggesting that systematic attention to error handling efficiency and return value processing could yield substantial performance improvements.

\subfile{../figures/code-changes/exception_regression}

\ul{API/Library Call Changes are among the most likely to introduce regressions, requiring careful cost-benefit evaluation despite moderate improvement potential.}
API/Library Call modifications demonstrate the highest regression rate among well-sampled categories at 30.3\% (101/333 method changes), resulting in an unfavorable risk ratio of 1.51, while showing only moderate improvement rates of 20.1\% (67/333 method changes). With 333 total commits representing the largest sample in our dataset, these statistics provide high statistical reliability. The elevated regression risk stems from unexpected behavioral differences between library implementations, version-specific performance characteristics, and the complexity of external dependencies that are difficult to predict during development. Figure~\ref{fig:cc-api-regression} demonstrates how seemingly minor API modernization efforts, such as replacing lambda expressions with method references, can result in measurable performance regressions (1.6\% slowdown) due to subtle differences in JVM optimization patterns.

\subfile{../figures/code-changes/api_regression}

\subfile{../figures/code-changes/api_improvement}

However, API modifications can also yield substantial benefits when properly implemented. Figure~\ref{fig:cc-api-improvement} demonstrates the positive potential of API changes, where replacing a simple \texttt{header()} call with the more sophisticated \texttt{appendHeader()} method using \texttt{Collections.singletonList()} achieved a remarkable 57\% performance improvement across 335K+ method invocations. This example illustrates that while API changes carry elevated regression risk, they also offer significant optimization opportunities when developers choose more efficient API alternatives.

\ul{Control Flow and Refactoring modifications exhibit balanced risk profiles, though refactoring introduces unexpected regression risks despite expectations of performance maintenance or improvement.}
Control Flow Changes and Refactoring \& Code Cleanup modifications demonstrate the most balanced performance impact profiles, with risk ratios of 1.13 and 1.06, respectively, which are the closest to unity among all categories. Control Flow changes achieve a 24.0\% improvement rate with a 27.0\% regression rate across 296 method changes. Figure~\ref{fig:cc-control-flow-improvement} demonstrates how introducing conditional guards can yield substantial performance gains, with a simple metrics collection condition achieving a 28\% improvement across 11M method executions. Conversely, Figure~\ref{fig:cc-control-flow-regression} illustrates how seemingly minor control flow reordering can introduce performance penalties, where moving Android JVM detection earlier in a decision tree resulted in a 22.8\% regression, highlighting the importance of understanding execution frequency patterns in conditional logic.

\subfile{../figures/code-changes/control_flow_improvement}

\subfile{../figures/code-changes/control_flow_regression}

Refactoring changes show 23.7\% improvements and 25.0\% regressions across 76 method changes. Counterintuitively, refactoring operations intended to improve code readability can introduce performance regressions, as illustrated in Figure~\ref{fig:cc-refactoring-regression} where a simple parameter rename from \texttt{supplier} to \texttt{statementSupplier} resulted in a 1.9\% performance regression across 1.85M+ method invocations, demonstrating how even minor code cleanup efforts can have measurable performance impacts due to subtle changes in bytecode generation or JVM optimization patterns. These balanced patterns suggest that developers have developed relatively reliable intuitions about the performance impact of structural code modifications, though the lower mean effect sizes (0.374 and 0.349, respectively) indicate that performance changes in these categories tend to be more modest in magnitude.

\subfile{../figures/code-changes/refactoring_regression}

\ul{Data Structure and Variable Changes represent the largest category with moderate but consistent performance impact potential, often yielding counterintuitive results that challenge theoretical expectations.}
Data Structure and Variable modifications constitute our largest sample with 408 method changes, achieving a 21.8\% improvement rate and 28.7\% regression rate with a balanced risk ratio of 1.31. While these changes show moderate effect sizes compared to algorithmic modifications, they demonstrate consistent performance impact patterns that often defy conventional wisdom about data type efficiency. Figure~\ref{fig:cc-data-structure-improvement} exemplifies this phenomenon through a particularly striking example where changing exponent storage from the smaller \texttt{byte} type to the larger \texttt{short} type in JSON parsing achieved a 20\% performance improvement across 275K+ method invocations.

\subfile{../figures/code-changes/data_structure_improvement}

This result directly contradicts theoretical expectations, as \texttt{byte} (8-bit) should theoretically outperform \texttt{short} (16-bit) due to reduced memory footprint and cache efficiency. However, our repeated benchmarking across multiple executions consistently demonstrated the same counterintuitive performance improvement. This unexpected outcome likely stems from complex JVM optimization behaviors, such as word-alignment preferences, reduced type conversion overhead, or improved JIT compiler optimization opportunities when working with 16-bit values that align with the processor architecture. The example simultaneously demonstrates how data structure modifications often serve dual purposes, where addressing functional requirements (handling larger exponent values $\ge256$) while inadvertently improving performance characteristics through mechanisms that are difficult to predict without empirical measurement. Such findings underscore the critical importance of performance-driven empirical validation over theoretical assumptions in modern JVM environments.

\ul{Concurrency modifications exhibit high-impact performance characteristics with extreme risk profiles.}
Concurrency changes demonstrate the most extreme performance pattern with a 50.0\% regression rate and 3.00 risk ratio, but this is based on only 6 method changes (3 regressions, 1 improvement, 2 unchanged), severely limiting statistical inference reliability. Despite the small sample, Figure~\ref{fig:cc-concurrency-improvement} illustrates the substantial performance potential of concurrency optimizations, where replacing synchronized blocks with lock-free \texttt{AtomicReferenceFieldUpdater.lazySet()} operations achieved a 3.6\% improvement across 513K+ method invocations. This pattern highlights both the critical importance of considering sample sizes when interpreting performance impact statistics and the high-impact nature of concurrency modifications when they do occur.

\subfile{../figures/code-changes/concurrency_improvement}

\begin{boxK}
    \textbf{RQ2 Conclusion and Implications}

    \ul{Performance Risk Assessment \& Development}: With no significant differences across change categories (p = 0.594), risk-stratified strategies based on code change type~\citep{kamei2013large,huang2014performance,kim2008classifying} are not justified. Teams should apply consistent performance validation across all modification types, focusing on implementation quality and context rather than categorical assumptions for method-level code modifications.

    \ul{Optimization Priority \& Resource Allocation}: Algorithmic Changes deliver the largest performance improvements (25.6\%) but carry substantial regression risk (33.9\%, risk ratio 1.33). Teams should pursue these high-reward improvements with robust automated testing, staged deployments, and clear rollback procedures.

    \ul{Challenging Intuition \& Broadening Scope}: Exception and Return Handling shows strong improvement potential (25.4\%, risk ratio 1.11), revealing that error-handling and return-logic paths represent underexplored optimization opportunities that developers should systematically consider.

    \ul{API Change Risk Management}: The highest regression rate in API/Library Call changes (30.3\%) necessitates enhanced validation protocols, including comprehensive performance testing across different library versions and implementation variants before deployment.

    \ul{Predictive Modeling \& Tool Development}: The negligible effect size between categories (Cramér's V = 0.057) indicates that change type classification provides minimal predictive value for performance outcomes. Future performance prediction models and developer tools should focus on implementation-specific factors rather than categorical change types for meaningful performance risk assessment.

    \ul{Testing Strategy \& Standardization}: As 20–40\% of changes exhibit large performance impacts uniformly across categories, teams should implement universal automated performance regression detection rather than category-specific testing protocols.
\end{boxK}

\subsection{RQ3: How do developer experience and code change complexity relate to performance impact magnitude?}

\noindent\textbf{Motivation}

Performance impact prediction remains a critical challenge in software development, with teams seeking reliable indicators to anticipate which code changes are likely to cause significant performance effects~\citep{kwon2015mantis,balsamo2004model,huang2010predicting}. Multiple characteristics of commits might have a correlation with performance variations, directly or indirectly. Here, we aim to investigate two key factors that are commonly assumed to influence performance outcomes: \textit{Developer Experience}, which reflects the expertise of a commit's author, and \textit{Code Change Complexity}, which quantifies the degree of change complexity introduced in a commit (i.e., for each method-level change individually).

Despite widespread acceptance of these assumptions in software engineering practice, the actual predictive power of developer experience and code change complexity for performance impact remains empirically unvalidated. Many development processes rely on these factors for resource allocation, code review prioritization, and testing strategies without rigorous evidence of their effectiveness. By systematically evaluating whether these commonly assumed predictors actually correlate with performance outcomes, we can either validate evidence-based development practices or reveal the limitations of traditional risk assessment approaches, leading to more effective performance management strategies.

\smallbreak

\noindent\textbf{Approach}

\underline{Code Change Complexity:} 
This metric is a fundamental factor influencing software maintainability, execution efficiency, and performance predictability. Changes that significantly alter the structure, scope, or size of a method can introduce performance overhead, whether through increased computational cost, memory usage, or control flow complexity. By quantifying the complexity of method-level changes, we aim to determine whether substantial performance variations are linked to specific types of complexity modifications, such as increased control flow statements, larger code blocks, or alterations in exception handling. The intuitive assumption is that complex modifications carry a higher risk of performance regressions due to their potential to disrupt established execution patterns, introduce inefficient algorithms, or create unexpected bottlenecks. Identifying this correlation can assist in developing early warning systems for performance regressions, optimizing refactoring efforts, and guiding best practices for managing complexity in performance-critical code sections. If complexity proves predictive, development teams could prioritize performance testing for high-complexity changes, allocate additional review time for intricate modifications, and implement automated risk assessment tools.

To quantify the complexity of method-level code changes, we introduce a structured scoring model that evaluates multiple dimensions of code modifications. Our approach builds upon methodologies established in prior research~\citep{kim2008classifying, hassan2009predicting, kamei2013large}, such as defining the code change complexity metrics and weights, and extends them by incorporating a finer-grained analysis of structural, scope-related, and size-based changes.

Given a pair of method versions (i.e., before and after) we parse the textual representation of their differences, extracted from version control diffs. The complexity of a change is determined based on three key dimensions: structural complexity, scope complexity, and size complexity, each of which is assigned a weighted score based on its impact on maintainability and cognitive load.

\begin{itemize}
\item \textit{Structural Complexity:} Structural modifications significantly impact code comprehensibility and maintainability. We identify changes to control flow constructs such as \texttt{if}, \texttt{else}, loops (\texttt{for}, \texttt{while}, \texttt{do-while}), and switch statements. Since changes in control structures introduce new execution paths and conditional branches, their complexity weight is set higher than basic modifications. Additionally, modifications to method signatures and exception handling mechanisms contribute to structural complexity, as they affect how methods interact within the system. Each control flow addition or modification receives a base weight of 2.5, method signature changes receive a weight of 2.0, and exception handling modifications receive a weight of 2.0, reflecting their relative impact on code understanding and maintenance effort. These weights are derived from empirically validated cognitive complexity research that uses neuroimaging and human comprehension studies to quantify the cognitive load of different programming constructs~\citep{wang2003cognitive,hassan2009predicting}. The Wang-Shao cognitive weight framework demonstrates that iterative structures (loops) require approximately 2.5 times more cognitive effort than baseline sequential code, while branching structures and exception handling patterns require 2.0 times more effort, as validated through controlled experiments measuring actual programmer comprehension performance~\citep{wang2003cognitive}.

\item \textit{Scope Complexity:} Scope-related changes introduce new variables or alter existing types, both of which can increase the difficulty of understanding and debugging the modified code. To quantify this, we track the introduction of new variable declarations and type modifications. Each newly introduced variable adds to the complexity score, reflecting the added cognitive load on developers who must comprehend its purpose and usage within the modified method. New variable declarations receive a weight of 1.5, while type modifications receive a weight of 1.5, as both represent significant changes to the method's data handling and variable management that require developers to understand new data relationships and type interactions.

\item \textit{Size Complexity:} The extent of code modification plays a fundamental role in determining change complexity. We measure the number of modified lines and assign a base weight per changed line (1.0 per line). Furthermore, continuous modifications in a localized region of the code, referred to as \textit{chunk size}, receive an additional complexity increment. This accounts for the increased effort required to review and validate large contiguous changes compared to scattered modifications. Chunks containing multiple continuous lines receive an additional multiplier of 1.2, reflecting the increased difficulty of comprehending large contiguous modifications that require developers to understand the combined effect of multiple related changes.

\end{itemize}

The total complexity score for a method change is computed as the sum of the weighted contributions from these three dimensions, along with a base complexity value (1.0) assigned to every change. By integrating structural, scope, and size factors, our model provides a comprehensive and adaptable measure of code change complexity: 

$Complexity = 1.0 + \sum (ControlFlow \times 2.5 + MethodSig \times 2.0 + Exception \times 2.0) + \sum (VarIntro \times 1.5 + TypeChange \times 1.5) + (LOCModified \times 1.0 + ChunkSize \times 1.2)$

While traditional measures such as lines of changed code (LOC) or cyclomatic complexity changes provide useful baselines, our weighted approach addresses their limitations for performance analysis. Simple LOC counts treat all code modifications equally, failing to distinguish between low-impact variable renames and high-impact algorithmic changes~\citep{wang2003cognitive}. Similarly, cyclomatic complexity focuses solely on control flow without considering data structure modifications or exception handling patterns that significantly influence performance~\citep{hassan2009predicting}. Our composite measure addresses these gaps by incorporating cognitive load research and empirically validated weights that reflect the differential impact of various change types on code comprehension and execution efficiency.

\smallbreak

\underline{Developer Experience:} 
This metric plays a crucial role in software evolution, influencing code quality, maintainability, and ultimately, performance. More experienced developers tend to have a deeper understanding of system architecture, efficient coding practices, and performance optimizations, whereas less experienced developers may introduce suboptimal changes, leading to unintended performance regressions. The prevailing assumption is that senior developers produce more performance-conscious code due to their accumulated knowledge of performance patterns, optimization techniques, and system-level understanding. By analyzing the relationship between developer experience and performance variations, we can uncover whether performance improvements are more frequently associated with seasoned contributors or if specific patterns of expertise contribute to high-impact optimizations. Understanding this correlation can help prioritize code reviews, refine mentoring strategies, and enhance automated performance testing by focusing on areas of the codebase where less experienced developers contribute changes that have high potential for performance degradation. If experience proves predictive, teams could implement experience-based review assignments, targeted mentoring programs, and risk-stratified testing protocols.

Understanding the impact of developer experience on code quality is a well-established area of research~\citep{fritz2007does, rahman2011ownership, dieste2018empirical}. In this study, we aimed to quantify the experience of authors who contributed to commits containing method changes. Following prior research on multi-dimensional developer experience quantification~\citep{kamei2013large,rahman2011ownership}, we designed a comprehensive scoring approach that integrates multiple indicators of experience, ensuring a holistic evaluation rather than relying on single metrics that may not capture the full spectrum of developer expertise.

To ensure temporal validity and avoid anachronistic bias, all developer experience metrics were collected up to the specific date of each commit under analysis. For instance, if a commit was made on January 15th, 2023, we calculated the developer's experience metrics based only on their activity history up until January 15th, 2023, not including any subsequent activity up to the date of our research. This temporal approach ensures that our experience measurements reflect the actual knowledge and expertise available to the developer at the time of making each specific commit, providing a more accurate and methodologically sound assessment of the relationship between experience and performance outcomes.

We chose four key metrics to assess developer experience, each capturing different aspects of their proficiency:

\begin{itemize}
    \item {GitHub account age ($Age$)}: A developer's account age, measured in years, provides an estimate of their overall exposure to software development activities and the GitHub ecosystem. While it does not directly indicate skill level, a longer account age suggests prolonged engagement with software development practices, version control workflows, and collaborative development processes, which may contribute to accumulated experience and familiarity with best practices.
    
    \item {Project-specific contributions ($Contrib_r$)}: This metric represents the total number of commits made by the developer within the specific project under analysis. A higher number of contributions indicates deeper familiarity with the project's codebase, architectural patterns, performance characteristics, and domain-specific requirements, making it a strong indicator of project-specific expertise. Developers with extensive project history are more likely to understand performance-critical sections, optimization opportunities, and potential regression risks.
    
    \item {Overall contributions across all projects ($Contrib_t$)}: Developers who contribute to multiple projects tend to acquire diverse coding experience, adapting to different codebases, programming styles, architectural patterns, and performance requirements. This metric captures the breadth of their experience across various repositories, programming languages, and development contexts. Exposure to diverse projects typically correlates with broader technical knowledge and adaptability to different performance optimization approaches.
    
    \item {Total code reviews performed ($Reviews$)}: Code review activities provide insights into a developer's understanding of software quality, best practices, performance considerations, and collaboration with peers. Developers who frequently participate in code reviews are likely to have a more refined understanding of software design principles, performance anti-patterns, optimization techniques, and maintainability concerns. Review experience also indicates leadership and mentoring capabilities within development teams.
\end{itemize}

Since the selected metrics have different scales and ranges (e.g., account age in years, contributions as raw counts), direct comparisons between them would be problematic. To ensure fair weighting and enable meaningful aggregation, we applied min-max normalization, transforming each metric to a range of [0,1].

After normalization, it was necessary to determine the relative importance of each metric in defining developer experience. Based on prior research findings and theoretical considerations about the relative importance of different experience indicators, we assigned empirically motivated weights to each metric. Project-specific contributions ($Contrib_r$) received the highest weight (0.3) as familiarity with the specific codebase is most directly relevant to performance-aware development within that project. Overall contributions across projects ($Contrib_t$) and code review experience ($Reviews$) each received equal weight (0.25), reflecting their importance for general development expertise and quality awareness. GitHub account age ($Age$) received the lowest weight (0.2) as it provides the least direct indication of active development skill, serving more as a general indicator of platform familiarity.

To compute a single unified experience score for each author, we first calculated the percentile ranks ($p_i$) of the normalized values $x_{norm}$. The percentile rank transformation ensures that scores are robust to outliers and better reflect relative experience levels across developers within the dataset. The final experience score $S$ was computed as the weighted sum of the percentile ranks:
\[
S = 0.3 \cdot p_{repo} + 0.25 \cdot p_{total} + 0.25 \cdot p_{reviews} + 0.2 \cdot p_{age}
\]
\noindent where $p_{repo}$, $p_{total}$, $p_{reviews}$, and $p_{age}$ represent the percentile ranks of project-specific contributions, total contributions, code reviews, and account age, respectively.

After computing experience scores, we categorized developers into distinct experience levels to enable categorical analysis and practical interpretation. Given the continuous nature of the scores, we employed K-Means clustering to partition developers into three groups: \textit{Junior}, \textit{Mid}, and \textit{Senior}. K-Means was chosen for its ability to identify natural groupings within data based on similarity rather than arbitrary threshold selection. By clustering developers according to their computed experience scores, we ensured that experience categories emerged from the data distribution rather than relying on predefined cutoffs that might not reflect the actual experience distribution in our dataset.

\begin{figure}[!b]
    \centering
    \includegraphics[width=0.9\columnwidth]{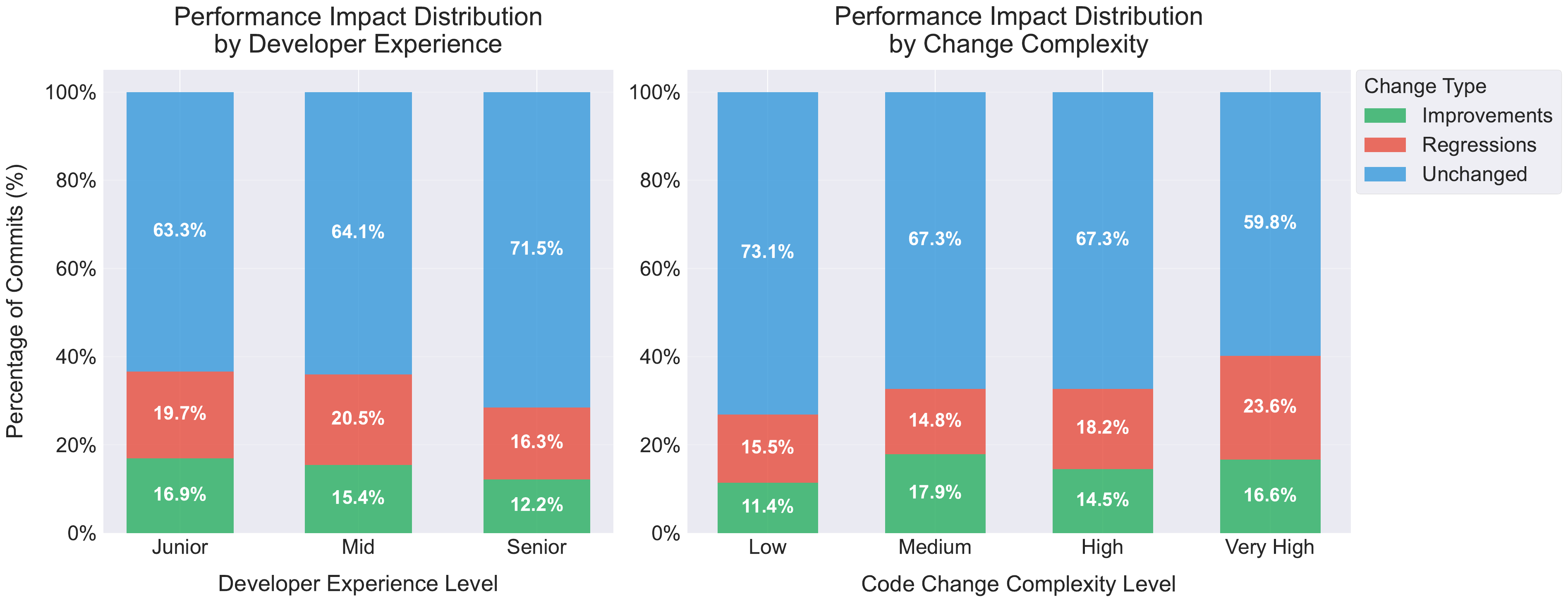}
    \captionsetup{justification=centering}
    \caption{Analysis of developer experience and code change complexity effects on performance impact.}
    \label{fig:rq3_complexity_impact}
\end{figure}

\subfile{../tables/rq3/experience_statistics}

\smallbreak

\noindent\textbf{Results}

\ul{Senior developers tend to make more stable changes, exhibiting both fewer performance improvements and fewer regressions compared to junior developers.}
Analysis of {\numMethodChanges} method changes across three experience categories reveals that senior developers create more stable performance outcomes overall. While junior developers cause performance improvements in 16.9\% of their commits [95\% CI: 13.1\%, 20.8\%] compared to only 12.2\% for senior developers [95\% CI: 9.8\%, 14.6\%], senior developers also produce fewer regressions, with regression rates decreasing from 19.7\% (Junior) to 16.3\% (Senior). This pattern indicates that senior developers make more conservative, stable changes rather than pursuing aggressive changes that could either significantly improve or degrade performance.

The stability effect is further evidenced by senior developers' lower performance impact magnitudes (mean absolute effect size: 0.159 vs 0.213 for Juniors), suggesting that when performance changes do occur, they tend to be more moderate. In practical terms, for every 100 commits, senior developers prevent 3-4 performance regressions compared to junior developers, but they also produce 4-5 fewer performance improvements. While these differences achieve statistical significance (Chi-square: $\chi^2$ = 11.05, p = 0.026; ANOVA: F = 8.99, p $<$ 0.001), the effect size is small (Cramér's V = 0.061), indicating that developer experience has a statistically detectable but practically modest influence on performance stability.

\subfile{../tables/rq3/complexity_statistics}

\ul{More complex code changes are more likely to bring significant performance changes, especially regressions, demonstrating increased performance instability with code modification complexity.}
Complex code changes exhibit a clear trend toward higher performance instability and regression risk. Very High complexity changes show the highest regression rate at 23.6\% compared to 15.5\% for Low complexity changes, representing an 8.1 percentage point increase in regression risk. This trend is consistent across the complexity spectrum: Low (15.5\%), Medium (14.8\%), High (18.2\%), and Very High (23.6\%) regression rates, demonstrating that complexity systematically increases the likelihood of performance degradation.

The instability pattern is further reinforced by the decreasing proportion of unchanged performance as complexity increases: Low complexity changes maintain stable performance in 73.1\% of cases, while Very High complexity changes preserve performance in only 59.8\% of cases. This 13.3 percentage point difference indicates that complex modifications are substantially more likely to impact performance in either direction. Interestingly, the relationship between complexity and performance improvements is non-linear: Medium complexity changes achieve the highest improvement rate (17.9\%), while Low complexity shows the lowest (11.4\%), suggesting an optimal complexity range for beneficial performance modifications.

While correlation analysis reveals a weak linear relationship between complexity scores and performance impact magnitude (Pearson r = 0.039, p = 0.134), the categorical analysis demonstrates statistical significance (Kruskal-Wallis H = 13.08, p = 0.004; Chi-square $\chi^2$ = 21.98, p = 0.001) with a small but meaningful effect size (Cramér's V = 0.086). This suggests that while complexity doesn't predict the magnitude of performance changes linearly, it reliably predicts the likelihood of performance instability, particularly regressions.

\ul{Traditional performance predictors demonstrate limited practical predictive power.} Traditional performance predictors in software engineering typically include complexity metrics such as cyclomatic complexity and Halstead metrics, along with code size measures and developer experience indicators~\citep{mccabe1976complexity,halstead1977elements,jay2009cyclomatic,hassan2009predicting}.
The pattern of statistically significant results with minimal effect sizes and limited explanatory power illustrates a critical challenge in software engineering research: large sample sizes can detect statistically significant differences that provide limited practical guidance for development processes. While both experience (p = 0.026) and complexity (p = 0.004) show statistically significant associations with performance impact derived from method-level code changes, neither provides robust, actionable predictive capability for development teams. The effect sizes (Cramér's V = 0.061 and 0.086) fall well below conventional thresholds for strong associations, indicating that the relationships, while statistically detectable, offer limited practical value for performance risk assessment.

\begin{boxK}
    \textbf{RQ3 Conclusion and Implications}
    
    \ul{For Development Team Practices}: The observed association between developer experience and code stability patterns suggests that development teams could benefit from strategically combining senior and junior developers in the development process. Since junior developers are more likely to be associated with aggressive changes while senior developers are associated with stable changes, teams could consider pairing approaches such as pair programming or developer-reviewer partnerships that leverage the complementary strengths of both experience levels.
    
    \ul{For Complexity-Based Testing and Risk Assessment}: The clear trend of increasing regression rates with complexity (15.5\% to 23.6\% from Low to Very High complexity) provides actionable guidance for testing prioritization and risk assessment. The 13.3 percentage point difference in stability between Low and Very High complexity changes (73.1\% vs 59.8\% unchanged performance) justifies implementing more rigorous performance testing protocols and validation procedures for high-complexity modifications while maintaining standard validation processes for simpler changes.
    
    \ul{For Alternative Research Directions}: Since commonly assumed predictors of developer experience and code complexity provide only modest explanatory power for performance outcomes, the field should investigate additional factors such as code context, architectural patterns, and domain-specific characteristics that may better predict performance impact magnitude and direction.
\end{boxK}

\subsection{RQ4: Are there significant differences in performance evolution patterns across different domains or project sizes?}

\noindent\textbf{Motivation}

Different project domains, such as System Programming, Data Processing, or Web Server, interact uniquely with hardware and system resources, potentially leading to varying degrees of performance impact. For example, system programming projects may be more sensitive to low-level optimizations and memory management, whereas data processing applications might be influenced by algorithmic efficiency and I/O performance. Web server applications face unique challenges related to concurrent request handling and network optimization. Similarly, project size can affect performance stability and the frequency of code changes, as larger codebases tend to experience more refactoring and maintenance updates that may introduce performance variability. Despite the critical nature of these factors, there is limited empirical understanding of how project domain and size correlate with performance evolution patterns. By investigating these relationships systematically, we aim to uncover actionable patterns that can inform domain-specific and size-appropriate performance management strategies.

\smallbreak

\noindent\textbf{Approach}

We employed two primary categorization approaches to assess project characteristics and their relationship with performance evolution patterns: domain classification and size measurement. This dual-factor approach enables a comprehensive analysis of how both functional application area and project scale influence performance change behaviors. By examining domain and size effects both independently and through interaction analysis, we can identify whether performance patterns are driven by application-specific requirements, project complexity, or their combined influence.

\ul{Domain Classification:} Project domains were determined by systematically analyzing each project's documentation, README files, and primary functionality, ensuring accurate classification based on intended use and application area. We identified six primary domains: System Programming, Data Processing, Networking, Web Server, Monitoring, and Testing. This classification captures the diversity of performance requirements across different application contexts, from low-level system operations to high-level web services, enabling domain-specific performance pattern analysis.

\ul{Project Size Measurement and Categorization:} Project size was measured using thousands of lines of code (KLOC), calculated as the total number of lines of code divided by 1,000. Based on the distribution of project sizes in our dataset, we established three size groups using quartile-based thresholds: Small ($\leq$33rd percentile), Medium (33rd-67th percentile), and Large ($>$67th percentile). These thresholds were chosen to ensure balanced representation across size groups while creating meaningful statistical divisions in the data distribution, enabling robust statistical comparisons across different project scales.

\begin{figure}
    \centering
    \includegraphics[width=\columnwidth]{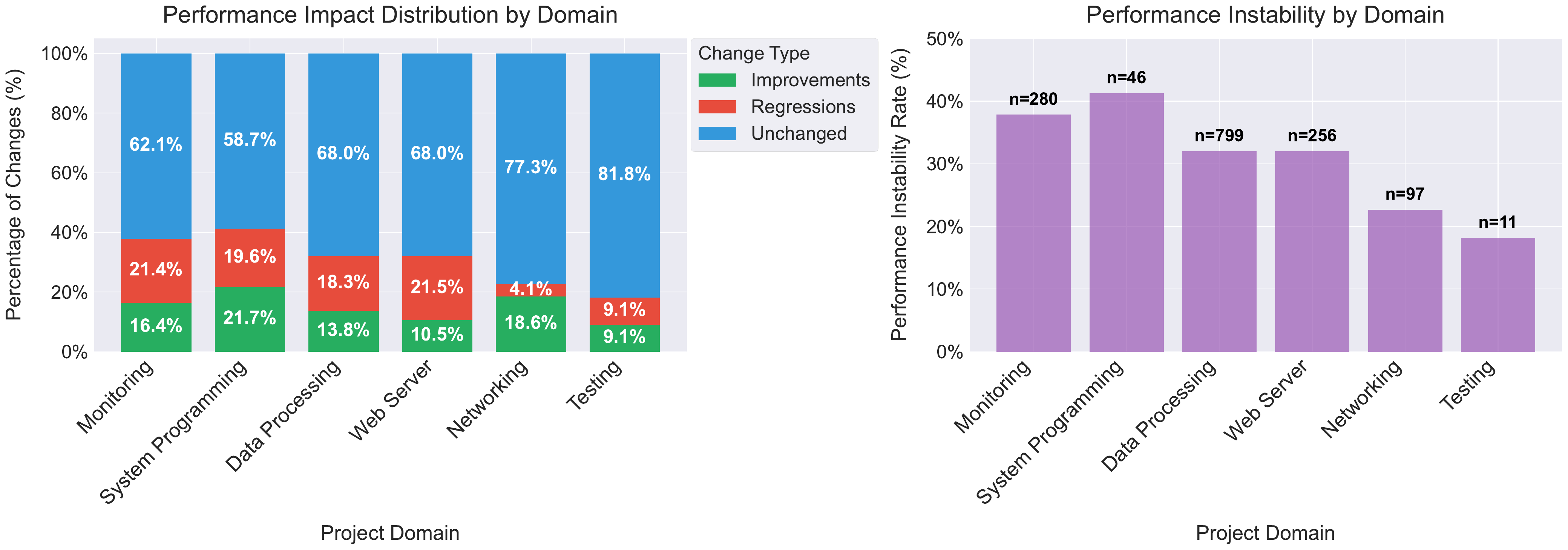}
    \captionsetup{justification=centering}
    \caption{Performance impact distribution and instability rates by project domain. The left panel shows the proportion of commits resulting in improvements, regressions, and unchanged performance for each domain. The right panel displays performance instability rates with sample sizes annotated. \textit{n} indicates the proportion of method-level code changes in each domain group.}
    \label{fig:rq4_domain_analysis}
\end{figure}

\begin{figure}[!b]
    \centering
    \includegraphics[width=\columnwidth]{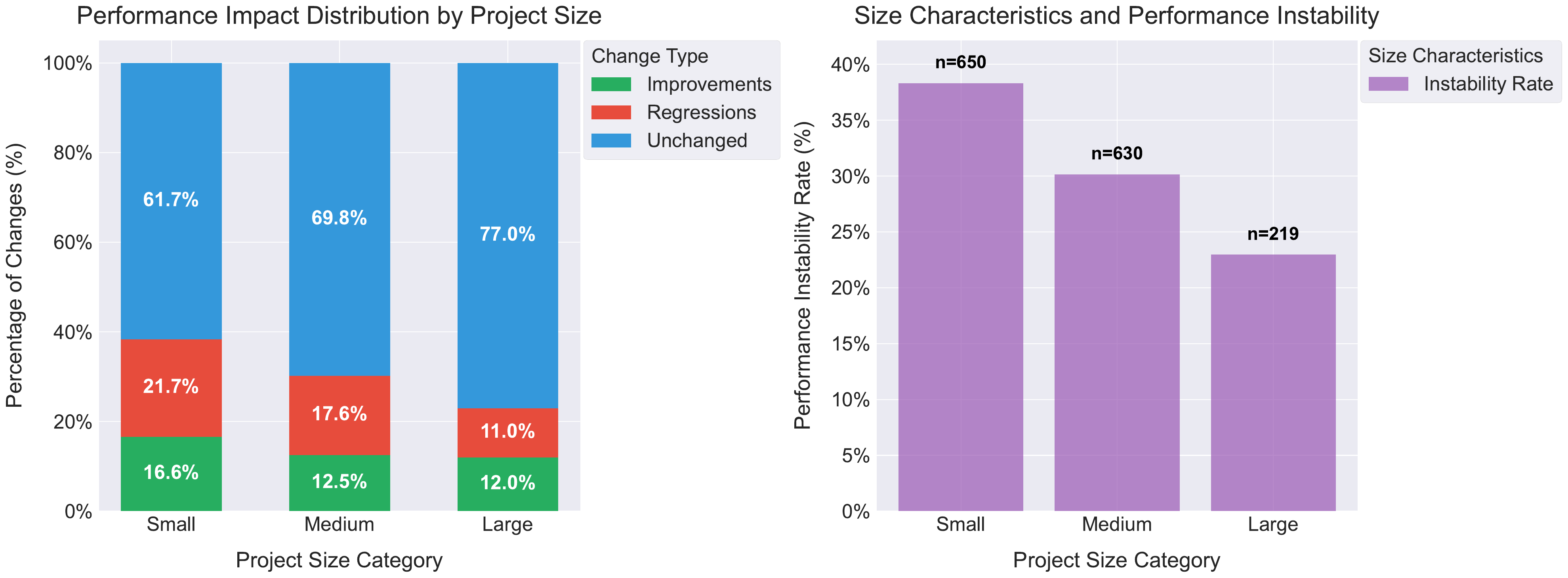}
    \captionsetup{justification=centering}
    \caption{Performance impact distribution by project size category. The left panel shows the proportional impact distribution across size categories. The right panel displays the relationship between project size characteristics and performance instability rates. \textit{n} indicates the proportion of method-level code changes in each size group.}
    \label{fig:rq4_size_analysis}
\end{figure}

\smallbreak

\noindent\textbf{Results}

\ul{Domain-specific performance patterns reveal significant variability in performance stability across different application areas.} Analysis of {\numMethodChanges} method changes across six domains reveals statistically significant differences in performance impact distributions ($\chi^2$ = 24.25, p = 0.007, Cramér's V = 0.090). Testing (n=11) shows the lowest instability rate at 18.2\%, though the very small sample size limits the generalizability of this finding. Among domains with substantial sample sizes, Networking (n=97) emerges as the most stable with 22.7\% instability, exhibiting an asymmetric pattern with relatively high improvement rates (18.6\%) but exceptionally low regression rates (4.1\%). This suggests that networking code changes are either carefully managed to avoid regressions or that the domain's performance characteristics make improvements more detectable than degradations.

Data Processing (n=799) and Web Server (n=256) domains demonstrate similar moderate stability with identical instability rates of 32.0\%. However, they differ significantly in impact magnitude: Data Processing shows moderate effect sizes with means of 0.350 ($\pm$0.244 $std$) for improvements and 0.368 ($\pm$0.240 $std$) for regressions, while Web Server domains exhibit substantially higher performance variability with mean absolute effect sizes of 0.603 ($\pm$0.343 $std$) for improvements and 0.593 ($\pm$0.331 $std$) for regressions. This indicates that while both domains have similar frequencies of performance changes, web server applications face fundamentally different performance challenges with more dramatic impacts, likely due to the complexity of concurrent request handling, network I/O optimization, and dynamic resource allocation requirements.

System Programming (n=46) demonstrates the highest instability rate at 41.3\% despite having relatively modest effect sizes (improvements: 0.265 $\pm$0.080, regressions: 0.298 $\pm$0.172). This pattern suggests that system-level modifications frequently impact performance but typically with moderate magnitudes, possibly reflecting the inherent sensitivity of low-level operations to code changes.

\begin{figure}
    \centering
    \includegraphics[width=\columnwidth]{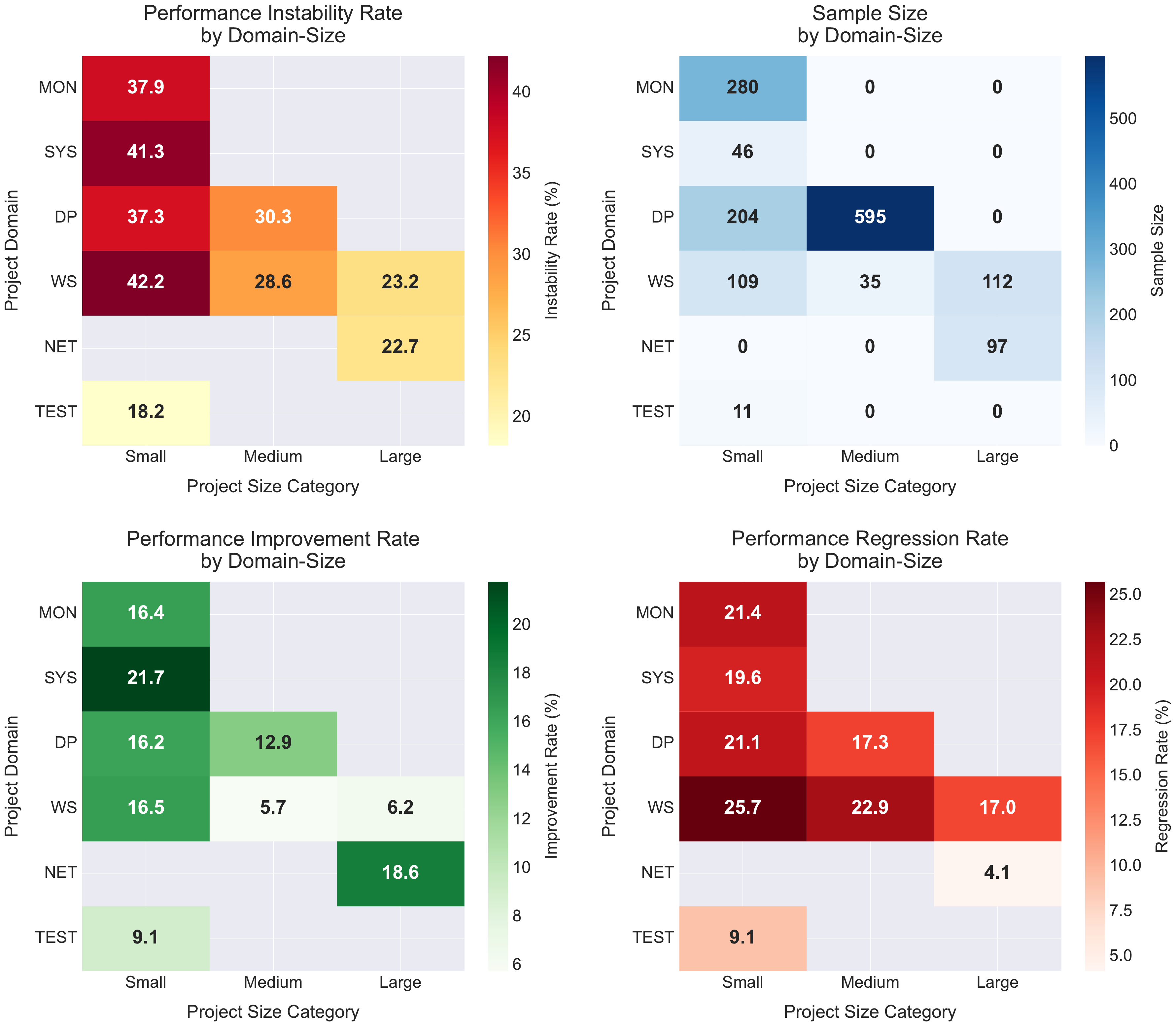}
    \captionsetup{justification=centering}
    \caption{Domain-size interaction patterns for performance impacts. The heatmaps show instability rates, sample sizes, improvement rates, and regression rates across all domain-size combinations.}
    \label{fig:rq4_interaction}
\end{figure}

\ul{Project size significantly influences performance impact patterns, with distinct non-linear effects across different project scales.} Chi-square analysis reveals significant differences in performance impact distributions across size categories ($\chi^2$ = 21.41, p $<$ 0.001, Cramér's V = 0.085). 
Small projects (n=650) demonstrate the highest performance instability at 38.3\%, with improvement rates of 16.6\% and regression rates of 21.7\%. These projects also exhibit high effect size variability, with mean absolute effect sizes of 0.431 ($\pm$0.300 $std$) for improvements and 0.478 ($\pm$0.316 $std$) for regressions. This high variability likely reflects the proportionally larger system-wide impacts that individual changes can have in smaller codebases.

Medium-sized projects (n=630) show intermediate stability with 30.2\% instability, 12.5\% improvement rates, and 17.6\% regression rates. They demonstrate more moderate and consistent effect sizes of 0.320 ($\pm$0.190 $std$) for improvements and 0.324 ($\pm$0.191 $std$) for regressions, with notably lower standard deviations indicating more predictable performance change magnitudes.

Although large projects (n=219) have the highest stability rate at 23.0\% instability and the lowest regression frequency at 11.0\%, the magnitudes of regressions are dramatically larger. While improvement effects remain moderate at 0.479 ($\pm$0.304 $std$), regression effects reach 0.637 ($\pm$0.320 $std$), representing 97\% higher regression impact compared to medium projects. This pattern indicates that while performance regressions are less frequent in large projects, when they do occur, they tend to be more severe, possibly due to complex interdependencies that amplify the negative effects of suboptimal changes.

\ul{Significant domain-size interaction effects reveal the need for tailored performance management strategies.} Two-way ANOVA confirms significant interaction effects between domain and size (F = 6.904, p $<$ 0.001), indicating that size effects vary substantially across different application domains. The most extreme combination is Web Server + Small projects, which exhibits 42.2\% instability with very high effect sizes (improvements: 0.638, regressions: 0.633). This represents the highest risk configuration in our analysis, likely because it combines the inherent performance complexity of web server applications (concurrent request handling, network I/O optimization) with the natural instability characteristics of smaller codebases where individual changes have disproportionate system-wide impacts. This combination requires intensive performance monitoring and validation.

Conversely, Web Server + Large projects show dramatically different characteristics with only 23.2\% instability, demonstrating that large web server projects achieve better performance stability. However, the regression effect size actually increases to 0.654, indicating that when performance problems do occur in large web applications, they tend to be more severe than in smaller web server projects.

The interaction patterns reveal domain-specific scaling behaviors that differ from overall trends. While the general pattern shows decreasing instability with project size (38.3\% → 30.2\% → 23.0\%), Web Server projects demonstrate this trend (42.2\% → 28.6\% → 23.2\%) but with an inverse relationship for regression severity, where larger projects experience more impactful regressions despite their lower frequency.

\ul{The observed domain and size patterns translate into concrete requirements for differentiated performance management approaches across software development contexts.} Based on the measured instability rates, System Programming domains exhibit the highest performance volatility at 41.3\% compared to Web Server domains at 32.0\%, indicating that system-level development teams require more intensive performance monitoring protocols. However, Web Server domains demonstrate significantly higher impact magnitudes when changes do occur, with effect sizes nearly double those of System Programming domains, necessitating different monitoring strategies focused on impact severity rather than frequency.

The asymmetric size effects reveal that development strategies must adapt to project scale based on underlying architectural and organizational factors. Small projects (38.3\% instability) experience high variability likely due to limited architectural constraints and the proportionally larger impact of individual changes on smaller codebases. Medium projects (30.2\% instability) benefit from established architectural patterns that provide stability while remaining manageable in complexity. Large projects (23.0\% instability) achieve lower change frequency through more rigorous development processes and architectural maturity, but when regressions do occur, they exhibit 96\% higher impact severity (effect size 0.637 vs 0.324 for medium projects), possibly due to complex interdependencies that amplify the effects of suboptimal changes across multiple system components.

This means that a performance regression in a large project is nearly twice as impactful as the same regression in a medium project, necessitating more conservative change management and comprehensive testing protocols as projects scale. Statistical validation across multiple approaches (Kruskal-Wallis H = 12.74, p = 0.026; Cohen's d = 0.246) confirms these differences are practically meaningful, not measurement noise.

\begin{boxK}
    \textbf{RQ4 Conclusion and Implications} 
    
    \ul{For Domain-Specific Performance Management}: Our results suggest that different domains exhibit distinct performance characteristics that may inform monitoring strategies. System Programming projects demonstrate the highest instability rates (41.3\%) but with relatively modest effect sizes, while Web Server domains show moderate instability (32.0\%) but dramatically higher impact magnitudes (effect sizes 0.603 vs 0.265). This suggests that system programming teams may benefit from frequent monitoring for change detection, while web server teams may need to focus on impact severity assessment and mitigation strategies due to the potential for more dramatic performance effects.
    
    \ul{For Project Size-Based Development Strategies}: The empirical patterns reveal that performance management approaches may need to account for project scale characteristics. Small projects exhibit the highest instability (38.3\%) and regression rates (21.7\%), medium projects show intermediate patterns (30.2\% instability, 17.6\% regressions), while large projects demonstrate the lowest frequency of changes (23.0\% instability, 11.0\% regressions) but with substantially higher regression impact severity (96\% higher effect sizes). These patterns suggest that different scales may benefit from tailored approaches: small projects may require tolerance for frequent variations, medium projects may benefit from leveraging their moderate regression rates for optimization efforts, and large projects may need to prioritize regression prevention given the severity of impacts when they occur.
    
    \ul{For Risk-Based Resource Allocation}: The significant interaction effect (p $<$ 0.001) between domain and size indicates that combined risk assessment approaches may be more effective than single-factor strategies. Our findings suggest that Web Server + Small projects (42.2\% instability) represent the highest risk configuration, while combinations like Networking + Large projects (22.7\% instability) may require less intensive monitoring. Resource allocation strategies could potentially be optimized by considering these empirically observed risk patterns.
    
    \ul{For Performance Monitoring Tool Design}: The domain-specific patterns observed in our study suggest that monitoring tools may benefit from adaptive thresholds. Given that System Programming shows high change frequency but moderate impacts, while Web Server applications show moderate frequency but high impacts, monitoring systems could potentially optimize detection accuracy by incorporating domain-aware sensitivity settings and alert prioritization based on these empirically observed characteristics.
\end{boxK}

%% file: tables/projects_details.tex
\begin{sidewaystable}
\centering
\caption{Studied projects information. Age shows the development timeline in years, KLOC refers to lines of code / 1,000.}
\label{tab:projects-details}
\renewcommand{\arraystretch}{1.4}
\resizebox{\textheight}{!}{
\begin{tabular}{lcrrrrrrrrr}
\hline
\textbf{Project} &
  \textbf{Domain} &
  \textbf{Commits} &
  \textbf{Age} &
  \textbf{Stars} &
  \textbf{KLOC} &
  \textbf{\begin{tabular}[c]{@{}c@{}}Commits with\\ Method Changes\end{tabular}} &
  \textbf{\begin{tabular}[c]{@{}c@{}}Commits with\\ Benchmarks\end{tabular}} &
  \textbf{\begin{tabular}[c]{@{}c@{}}Candidate\\ Commits\end{tabular}} &
  \textbf{\begin{tabular}[c]{@{}c@{}}Executed\\ Commits\end{tabular}} &
  \textbf{\begin{tabular}[c]{@{}c@{}}Collected\\ Method Changes\end{tabular}} \\\hline
eclipse/jetty.project        & Web Server         & 30,160 & 13.5 & 3,858  & 339.06 & 2,472 & 12,720 & 2,472 & 56  & 124 \\
netty/netty                  & Networking         & 11,604 & 14.2 & 34,850 & 216.98 & 4,241 & 7,669  & 4,240 & 57  & 97  \\
jdbi/jdbi                    & Data Processing    & 5,709  & 15.6 & 1,984  & 28.49  & 1,266 & 1,919  & 313   & 90  & 136 \\
alibaba/fastjson2            & Data Processing    & 4,372  & 2.8  & 3,769  & 178.5  & 1,726 & 3,725  & 1,726 & 220 & 615 \\
OpenHFT/Chronicle-Core       & System Programming & 3,911  & 9.9  & 576    & 13.25  & 780   & 3,170  & 585   & 2   & 3   \\
arnaudroger/SimpleFlatMapper & Data Processing    & 3,433  & 10.5 & 440    & 51.79  & 911   & 1,969  & 485   & 45  & 68  \\
elastic/apm-agent-java       & Monitoring         & 3,066  & 6.9  & 567    & 80.22  & 891   & 2,984  & 889   & 86  & 176 \\
openzipkin/zipkin            & Monitoring         & 2,955  & 12.6 & 16,982 & 23.51  & 656   & 2,726  & 615   & 46  & 93  \\
OpenFeign/feign              & Web Server         & 2,063  & 11.6 & 9,430  & 17.42  & 351   & 1,384  & 229   & 54  & 114 \\
protostuff/protostuff        & Data Processing    & 1,603  & 11.0 & 2,050  & 42.29  & 448   & 1,354  & 448   & 4   & 4   \\
JCTools/JCTools              & System Programming & 1,043  & 11.2 & 3,572  & 31.48  & 339   & 1,042  & 339   & 26  & 52  \\
easymock/objenesis           & Testing            & 1,049  & 11.5 & 604    & 2.69   & 107   & 784    & 72    & 12  & 14  \\
prometheus/client\_java      & Monitoring         & 866    & 12.0 & 2,155  & 27.38  & 155   & 667    & 154   & 9   & 11  \\
eclipse-ee4j/jersey          & Web Server         & 798    & 6.8  & 692    & 158.91 & 310   & 781    & 319   & 23  & 37  \\
HdrHistogram/HdrHistogram    & Monitoring         & 779    & 12.4 & 2,173  & 8.89   & 158   & 317    & 73    & 9   & 15 
\end{tabular}%
}
\end{sidewaystable}

%% file: tables/rq1/overall_distribution.tex
\begin{table}[!b]
\centering
\captionsetup{justification=centering}
\caption{Overall Performance Impact Distribution}
\label{tab:rq1_performance_distribution}
\renewcommand{\arraystretch}{1.3}
\begin{tabular}{lrr}
\textbf{Performance Change Type} & \textbf{Count} & \textbf{Percentage} \\
\hline
Improvements & 212 & 14.2\% \\
Regressions & 275 & 18.5\% \\
Unchanged & 1,002 & 67.3\% \\
\cline{2-3}
\textbf{Total} & \textbf{1,499} & \textbf{100\%} \\
\end{tabular}
\end{table}

%% file: tables/rq2/code_change_categories.tex
\begin{sidewaystable}
\centering
\renewcommand{\arraystretch}{1.4}
\caption{Code change categories used for performance impact analysis.}
\label{tab:code-change-categories}
\resizebox{\textheight}{!}{
\begin{tabular}{lll}
\textbf{ID} & \textbf{Category} & \textbf{Description} \\
\hline
ALG & Algorithmic Change~\cite{barry1999detailed,zhao2022large} & Changes in processing logic, methodology, or computational approach \\
CF & Control Flow~\cite{barry1999detailed,fluri2006classifying} & Modifications to control structures (if, while, for, switch, loops) \\
DS & Data Structure \& Variable~\cite{barry1999detailed,zhao2022large} & Changes to variable types, names, values, or data structures \\
REF & Refactoring \& Code Cleanup~\cite{fowler2018refactoring,chapin2001types} & Code simplification and restructuring without functional changes \\
EH & Exception \& Return Handling$^*$ & Changes in exception handling, error management, or return statements \\
CON & Concurrency~\cite{zhao2022large,abbaspour2017concurrency} & Introduction of threading, synchronization, or asynchronous operations \\
API & API/Library Call~\cite{barry1999detailed,sanchez2020tandem} & Modifications to external API calls, library usage, or dependencies \\
\end{tabular}
}
\\
\vspace{2mm}
\small
{Note: Categories are not mutually exclusive; changes exhibiting characteristics of multiple categories were assigned all relevant labels to ensure comprehensive representation of modification complexity. $^*$ indicates categories refined or introduced in this study.}
\end{sidewaystable}

%% file: tables/rq2/code_change_type_proportions.tex
\begin{table}
\renewcommand{\arraystretch}{1.5}
\captionsetup{justification=centering}
\caption{Performance impact analysis by code change category. Risk is calculated as the regression-to-improvement ratio.\\
ES = Effect Size (mean absolute effect size for improvements and regressions only, excluding unchanged performance).}
\label{tab:code-change-proportion}
\resizebox{\textwidth}{!}{
\begin{tabular}{lcccccc}
\textbf{Category} & \textbf{Total} & \textbf{Improvements} & \textbf{Regressions} & \textbf{Unchanged} & \textbf{Risk} & \textbf{Mean $|$ES$|$} \\
\hline
Data Structure \& Variable & 408 & 89 (21.8\%) & 117 (28.7\%) & 202 (49.5\%) & 1.31 & 0.410 \\
API/Library Call & 333 & 67 (20.1\%) & 101 (30.3\%) & 165 (49.5\%) & 1.51 & 0.446 \\
Control Flow & 296 & 71 (24.0\%) & 80 (27.0\%) & 145 (49.0\%) & 1.13 & 0.374 \\
Algorithmic Change & 227 & 58 (25.6\%) & 77 (33.9\%) & 92 (40.5\%) & 1.33 & 0.470 \\
Exception \& Return Handling & 213 & 54 (25.4\%) & 60 (28.2\%) & 99 (46.5\%) & 1.11 & 0.476 \\
Refactoring \& Code Cleanup & 76 & 18 (23.7\%) & 19 (25.0\%) & 39 (51.3\%) & 1.06 & 0.349 \\
Concurrency & 6 & 1 (16.7\%) & 3 (50.0\%) & 2 (33.3\%) & 3.00 & 0.427 \\
\end{tabular}
}
\end{table}

%% file: figures/code-changes/algorithm_improvement.tex
\begin{figure}[]
\begin{minipage}{0.48\textwidth}
\begin{lstlisting}[style=javastyle]
// Before
public HttpCall<Void> build() {
    // ... previous code ...
    @r@ByteBufAllocator alloc = RequestContext.mapCurrent(@r@
        @r@RequestContext::alloc, () -> PooledByteBufAllocator.DEFAULT);@r@
    @r@HttpCall.RequestSupplier request = new BulkRequestSupplier(@r@
        @r@entries, shouldAddType, RequestHeaders.of(HttpMethod.POST, @r@
        @r@urlBuilder.toString(), HttpHeaderNames.CONTENT_TYPE, @r@
        @r@MediaType.JSON_UTF_8), alloc);@r@
    return http.newCall(request, CHECK_FOR_ERRORS, tag);
}
\end{lstlisting}
\end{minipage}
\hfill
\begin{minipage}{0.48\textwidth}
\begin{lstlisting}[style=javastyle]
// After
public HttpCall<Void> build() {
    // ... previous code ...
    @g@final HttpData body;@g@
    @g@CompositeByteBuf sink = RequestContext.mapCurrent(@g@
        @g@RequestContext::alloc, () -> PooledByteBufAllocator.DEFAULT)@g@
        @g@.compositeHeapBuffer(Integer.MAX_VALUE);@g@
    @g@try {@g@
        @g@for (IndexEntry<?> entry : entries) {@g@
            @g@write(sink, entry, shouldAddType);@g@
        @g@}@g@
        @g@body = HttpData.wrap(ByteBufUtil.getBytes(sink));@g@
    @g@} finally {@g@
        @g@sink.release();@g@
    @g@}@g@
    @g@AggregatedHttpRequest request = AggregatedHttpRequest.of(@g@
        @g@RequestHeaders.of(HttpMethod.POST, urlBuilder.toString(), @g@
        @g@HttpHeaderNames.CONTENT_TYPE, MediaType.JSON_UTF_8), body);@g@
    return http.newCall(request, CHECK_FOR_ERRORS, tag);
}
\end{lstlisting}
\end{minipage}
\caption{Algorithmic optimization in the \texttt{build()} method from \textit{zipkin} (commit $8423afc$). The change replaces the lazy request supplier pattern with direct buffer allocation and serialization using \texttt{CompositeByteBuf} and try-finally resource management. This algorithmic improvement achieved an $85\%$ performance improvement across $97$K+ method invocations.}
\label{fig:cc-algorithmic-improvement}
\end{figure}

%% file: figures/code-changes/exception_regression.tex
\begin{figure}[]
\begin{minipage}{0.4\textwidth}
\begin{lstlisting}[style=javastyle]
// Before
@Nullable
protected HttpURLConnection startRequest(String endpoint) throws IOException {
    final HttpURLConnection connection = apmServerClient.startRequest(endpoint);
    if (connection != null) {
        // ... connection logic ...
    }
    return connection;
}
\end{lstlisting}
\end{minipage}
\hfill
\begin{minipage}{0.58\textwidth}
\begin{lstlisting}[style=javastyle]
// After
@Nullable
protected HttpURLConnection startRequest(String endpoint) throws IOException {
    final HttpURLConnection connection = apmServerClient.startRequest(endpoint);
    if (connection != null) {
        @g@try {@g@
            // ... connection logic ...
        @g@} catch (IOException e) {@g@
            @g@logger.error("...");@g@
            @g@if (logger.isInfoEnabled() && connection instanceof HttpsURLConnection) {@g@
                @g@HttpsURLConnection httpsURLConnection = (HttpsURLConnection) connection;@g@
                @g@// ... Log SSL debug info: cipher suites, certificates, etc. ...@g@
                @g@// ... Multiple try-catch blocks for SSL properties ...@g@
            @g@}@g@
            @g@throw e;@g@
        @g@}@g@
    }
    return connection;
}
\end{lstlisting}
\end{minipage}
\caption{Exception handling enhancement with SSL debug information in the \texttt{startRequest()} method from \textit{apm-agent-java} (commit $379519e$). The change wraps the connection setup in a try-catch block and adds comprehensive SSL debugging for \texttt{HttpsURLConnection} instances when errors occur. This debugging enhancement resulted in a $3.1\%$ performance regression across $275$ method invocations, demonstrating the performance cost of comprehensive error handling and logging.}
\label{fig:cc-exception-regression}
\end{figure}

%% file: figures/code-changes/api_regression.tex
\begin{figure}[]
\begin{minipage}{0.51\textwidth}
\begin{lstlisting}[style=javastyle]
// Before
@Override
public Optional<ColumnMapper<?>> build(QualifiedType<?> givenType, ConfigRegistry config) {
    return Optional.of(givenType.getType())
        .map(GenericTypes::getErasedType)
        .@r@filter(c -> Enum.class.isAssignableFrom(c))@r@
        .map(clazz -> makeEnumArgument(
            (QualifiedType<Enum>) givenType, 
            (Class<Enum>) clazz, 
            config));
}
\end{lstlisting}
\end{minipage}
\hfill
\begin{minipage}{0.46\textwidth}
\begin{lstlisting}[style=javastyle]
// After
@Override
public Optional<ColumnMapper<?>> build(QualifiedType<?> givenType, ConfigRegistry config) {
    return Optional.of(givenType.getType())
        .map(GenericTypes::getErasedType)
        .@g@filter(Enum.class::isAssignableFrom)@g@
        .map(clazz -> makeEnumArgument(
            (QualifiedType<Enum>) givenType, 
            (Class<Enum>) clazz, 
            config));
}
\end{lstlisting}
\end{minipage}
\caption{Java API modernization in the \texttt{build()} method from \textit{jdbi} (commit $fd7458f$). The change replaces a lambda expression with a method reference for enum type filtering. Despite the modest $1.6\%$ performance regression, this change is statistically significant across $3.9$M+ method invocations (p-value $< 0.001$, effect size $= -0.21$), demonstrating that even small code modernization efforts can have measurable performance impacts at scale.}
\label{fig:cc-api-regression}
\end{figure}

%% file: figures/code-changes/api_improvement.tex
\begin{figure}[]
\begin{minipage}{0.48\textwidth}
\begin{lstlisting}[style=javastyle]
// Before
public RequestTemplate resolve(Map<String, ?> variables) {
    // ... header processing ...
    if (!header.isEmpty()) {
        resolved.@r@header@r@(headerTemplate.getName(), @r@header@r@);
    }
    // ... rest of method ...
}
\end{lstlisting}
\end{minipage}
\hfill
\begin{minipage}{0.48\textwidth}
\begin{lstlisting}[style=javastyle]
// After
public RequestTemplate resolve(Map<String, ?> variables) {
    // ... header processing ...
    if (!header.isEmpty()) {
        resolved.@g@appendHeader@g@(headerTemplate.getName(),
            @g@Collections.singletonList(header)@g@,
            @g@true@g@);
    }
    // ... rest of method ...
}
\end{lstlisting}
\end{minipage}
\caption{API call modification in the \texttt{resolve()} method from \textit{feign} (commit $92b2f51$). The change replaces \texttt{header()} call with \texttt{appendHeader()} using \texttt{Collections.singletonList()} for enhanced header handling. This API enhancement achieved a $57\%$ performance improvement across $335$K+ method invocations.}
\label{fig:cc-api-improvement}
\end{figure}

%% file: figures/code-changes/control_flow_improvement.tex
\begin{figure}[]
\begin{minipage}{0.48\textwidth}
\begin{lstlisting}[style=javastyle]
// Before
@Override
protected void afterEnd() {
    trackMetrics();
    this.tracer.endTransaction(this);
}
\end{lstlisting}
\end{minipage}
\hfill
\begin{minipage}{0.48\textwidth}
\begin{lstlisting}[style=javastyle]
// After  
@Override
protected void afterEnd() {
    @g@if (collectBreakdownMetrics) {@g@
        @g@trackMetrics();@g@
    @g@}@g@
    this.tracer.endTransaction(this);
}
\end{lstlisting}
\end{minipage}
\caption{Significant performance improvement in the \texttt{Transaction.afterEnd()} method from \textit{apm-agent-java} (commit $044b068$). The new version introduces a control flow modification, achieving a $28\%$ performance improvement across $\sim11$M method executions.}
\label{fig:cc-control-flow-improvement}
\end{figure}

%% file: figures/code-changes/control_flow_regression.tex
\begin{figure}[!b]
\begin{minipage}{0.48\textwidth}
\begin{lstlisting}[style=javastyle]
// Before
public <T> ObjectInstantiator<T> newInstantiatorOf(Class<T> type) {
    if (PlatformDescription.isThisJVM(HOTSPOT) || 
        PlatformDescription.isThisJVM(OPENJDK)) {
        // ... logic ...
    } else if (PlatformDescription.isThisJVM(JROCKIT)) {
        // ... logic ...
    @r@} else if (PlatformDescription.isThisJVM(DALVIK)) {@r@
        @r@// ... logic ...@r@
    }
    // ... rest of code ...
}
\end{lstlisting}
\end{minipage}
\hfill
\begin{minipage}{0.48\textwidth}
\begin{lstlisting}[style=javastyle]
// After
public <T> ObjectInstantiator<T> newInstantiatorOf(Class<T> type) {
    if (PlatformDescription.isThisJVM(HOTSPOT) || 
        PlatformDescription.isThisJVM(OPENJDK)) {
        // ... logic ...
    @g@} else if (PlatformDescription.isThisJVM(DALVIK)) {@g@
        @g@// ... logic ...@g@
    } else if (PlatformDescription.isThisJVM(JROCKIT)) {
        // ... logic ...
    }
    // ... rest of code ...
}
\end{lstlisting}
\end{minipage}
\caption{Control flow reorganization in the \texttt{newInstantiatorOf()} method from \textit{objenesis} (commit $dcceccb$). The change moves Android (DALVIK) JVM detection from third to second position in the decision tree, placing it before JRockit evaluation. This reordering resulted in a $22.8\%$ performance regression across $\sim100$ method invocations.}
\label{fig:cc-control-flow-regression}
\end{figure}

%% file: figures/code-changes/refactoring_regression.tex
\begin{figure}[]
\begin{minipage}{0.48\textwidth}
\begin{lstlisting}[style=javastyle]
// Before
public static ResultProducer<ResultBearing> returningResults() {
    return (@r@supplier@r@, ctx) -> 
        ResultBearing.of(getResultSet(@r@supplier@r@, ctx), ctx);
}
\end{lstlisting}
\end{minipage}
\hfill
\begin{minipage}{0.48\textwidth}
\begin{lstlisting}[style=javastyle]
// After
public static ResultProducer<ResultBearing> returningResults() {
    return (@g@statementSupplier@g@, ctx) -> 
        ResultBearing.of(getResultSet(@g@statementSupplier@g@, ctx), ctx);
}
\end{lstlisting}
\end{minipage}\caption{Parameter rename refactoring in the \texttt{returningResults()} method from \textit{jdbi} (commit $1f00a35$). It modifies the parameter name from \texttt{supplier} to \texttt{statementSupplier} for improved readability. This code cleanup resulted in a $1.9\%$ performance regression across $1.85$M+ method invocations, illustrating how even minor refactoring changes can have measurable performance impacts.}
\label{fig:cc-refactoring-regression}
\end{figure}

%% file: figures/code-changes/data_structure_improvement.tex
\begin{figure}[!b]
\begin{minipage}{0.48\textwidth}
\begin{lstlisting}[style=javastyle]
// Before
@Override
public float readFloatValue() {
    // ... parsing logic ...
    this.exponent = (@r@byte@r@) expValue;
    valueType = JSON_TYPE_DEC;
    // ... rest of method ...
}
\end{lstlisting}
\end{minipage}
\hfill
\begin{minipage}{0.48\textwidth}
\begin{lstlisting}[style=javastyle]
// After
@Override
public float readFloatValue() {
    // ... parsing logic ...
    this.exponent = (@g@short@g@) expValue;
    valueType = JSON_TYPE_DEC;
    // ... rest of method ...
}
\end{lstlisting}
\end{minipage}
\caption{Data type modification in the \texttt{readFloatValue()} method from \textit{fastjson2} (commit $39a221b$). The change replaces the \texttt{byte} type cast with \texttt{short} for exponent storage to handle larger exponent values ($\ge256$). This modification resulted in a $\sim20\%$ performance improvement across $275$K+ method invocations, fixing a bug while, perhaps unintentionally, enhancing JSON parsing performance.}
\label{fig:cc-data-structure-improvement}
\end{figure}

%% file: figures/code-changes/concurrency_improvement.tex
\begin{figure}[]
\begin{minipage}{0.44\textwidth}
\begin{lstlisting}[style=javastyle]
// Before
static void releaseCharArray(int cacheIndex, char[] chars) {
    // ... control logic ...
    @r@synchronized (cacheItem) {@r@
        @r@cacheItem.chars = chars;@r@
    @r@}@r@
}
\end{lstlisting}
\end{minipage}
\hfill
\begin{minipage}{0.54\textwidth}
\begin{lstlisting}[style=javastyle]
// After
@g@static final AtomicReferenceFieldUpdater<CacheItem, char[]> CHARS_UPDATER = AtomicReferenceFieldUpdater.newUpdater(CacheItem.class, char[].class, "chars");@g@
static void releaseCharArray(int cacheIndex, char[] chars) {
    // ... control logic ...
    @g@CHARS_UPDATER.lazySet(cacheItem, chars);@g@
}
\end{lstlisting}
\end{minipage}
\caption{Concurrency enhancement in the \texttt{releaseCharArray()} method from \textit{fastjson2} (commit $b46ccee$). The change replaces synchronized block with lock-free \texttt{lazySet()} operation and expands cache support. This concurrency optimization achieved a $3.6\%$ performance improvement across $513$K+ method invocations.}
\label{fig:cc-concurrency-improvement}
\end{figure}

%% file: tables/rq3/experience_statistics.tex
\begin{table}[]
\centering
\captionsetup{justification=centering}
\caption{Developer experience vs performance impact distribution with statistical significance tests.\\ ES = Effect Size. CI = Confidence Interval ([lower bound, upper bound]).}
\label{tab:rq3_experience_stats}
\renewcommand{\arraystretch}{1.5}
\resizebox{\textwidth}{!}{
\begin{tabular}{lrrrrrr}
\hline
\textbf{Experience} & \textbf{Total} & \textbf{Improvements} & \textbf{Regressions} & \textbf{Unchanged} & \textbf{Mean $|$ES$|$} & \textbf{95\% CI} \\
\hline
Junior & 360 & 61 (16.9\%) & 71 (19.7\%) & 228 (63.3\%) & 0.213 & [13.1\%, 20.8\%] \\
Mid & 409 & 63 (15.4\%) & 84 (20.5\%) & 262 (64.1\%) & 0.206 & [11.9\%, 18.9\%] \\
Senior & 730 & 90 (12.2\%) & 120 (16.3\%) & 520 (71.5\%) & 0.159 & [9.8\%, 14.6\%] \\
\end{tabular}
}
\end{table}

%% file: tables/rq3/complexity_statistics.tex
\begin{table}[]
\centering
\captionsetup{justification=centering}
\caption{Code change complexity vs performance impact with correlation analysis. ES = Effect Size.}
\label{tab:rq3_complexity_stats}
\renewcommand{\arraystretch}{1.3}
\resizebox{\textwidth}{!}{
\begin{tabular}{lrrrrr}
\hline
\textbf{Complexity} & \textbf{Total} & \textbf{Improvements} & \textbf{Regressions} & \textbf{Unchanged} & \textbf{Mean $|$ES$|$} \\
\hline
Low & 569 & 65 (11.4\%) & 88 (15.5\%) & 416 (73.1\%) & 0.154 \\
Medium & 196 & 35 (17.9\%) & 29 (14.8\%) & 132 (67.3\%) & 0.178 \\
High & 360 & 52 (14.5\%) & 65 (18.2\%) & 243 (67.3\%) & 0.182 \\
Very High & 374 & 62 (16.6\%) & 88 (23.6\%) & 224 (59.8\%) & 0.231 \\
\end{tabular}
}
\end{table}

%% file: sections/discussion.tex
\ul{Prevalence of Performance Variations Necessitates Systematic Continuous Integration:} 
We recommend that development teams integrate automated performance regression detection into their continuous integration pipelines as a standard practice, not an optional enhancement. Based on our analysis of {\numMethodChanges} method changes in RQ1, we observe that 32.7\% of method-level code changes result in measurable performance impacts, with regressions occurring 1.3 times more frequently than improvements. This suggests that performance variations are not exceptional events but routine occurrences in software development that require systematic monitoring. Developers could leverage this observation by implementing lightweight microbenchmarking frameworks (such as JMH for Java projects) that automatically execute performance tests for modified methods during the CI process, enabling early detection of performance regressions before they reach production environments.

\ul{Balanced Performance Monitoring Across All Code Change Categories:}
We highlight that risk-stratified performance testing strategies based solely on code change types provide limited practical value. Based on our analysis of code change categories in RQ2, we observe that no significant differences exist in performance impact distributions across modification types (p = 0.594, Cramér's V = 0.057), with all categories showing 20-40\% of changes producing large performance effects. This suggests that implementation quality and contextual factors outweigh categorical change classifications in determining performance outcomes. Developers could leverage this observation by applying consistent performance validation protocols across all code modifications rather than allocating testing resources based on perceived change type risk, ensuring comprehensive coverage of potential performance impacts.

\ul{Strategic Team Composition and Resource Allocation Based on Empirical Risk Factors:}
We recommend that development teams adopt differentiated performance testing strategies based on empirically validated risk factors rather than intuitive assumptions. Based on our analysis in RQ3, we observe that code change complexity correlates with regression likelihood (increasing from 15.5\% to 23.6\% across complexity levels), while developer experience influences change stability patterns. This suggests that testing resource allocation can be optimized through quantifiable risk assessment. Development teams could leverage these observations by implementing automated complexity scoring for code changes, allocating additional performance testing resources to high-complexity modifications, and establishing mentoring partnerships between senior and junior developers to balance aggressive optimization attempts with stability requirements.

\ul{Domain and Project Scale-Aware Performance Management:}
We suggest that organizations implement adaptive performance monitoring strategies tailored to their specific domain and project characteristics. Based on our analysis in RQ4, we observe significant interaction effects between project domain and size (F = 6.904, p $<$ 0.001), with Web Server + Small projects exhibiting 42.2\% instability while System Programming projects show 41.3\% instability but with lower impact magnitudes. This suggests that one-size-fits-all performance management approaches are suboptimal for diverse development contexts. Organizations could leverage these patterns by configuring domain-specific monitoring thresholds, implementing more frequent testing cycles for high-risk combinations (such as small web server projects), and establishing size-appropriate regression impact tolerances based on empirically observed severity patterns.

\ul{Contextual Performance Trade-off Documentation and Decision Support:}
We recommend establishing systematic documentation practices for performance-affecting changes that distinguish between intentional trade-offs and unintended regressions. Based on our observation that performance changes often involve trade-offs between execution efficiency and other quality attributes (security, maintainability, debugging capabilities), development teams require better decision support frameworks. This suggests that performance management should extend beyond simple regression prevention to include trade-off transparency and impact assessment. Developers could implement structured performance change documentation protocols that capture the rationale for performance-affecting modifications, enabling informed decision-making about acceptable performance costs for functional or security enhancements, and providing historical context for future optimization efforts.

\ul{Automated Performance Analysis Integration for Proactive Development:}
We highlight the potential for integrating automated performance analysis tools throughout the development lifecycle to provide continuous feedback on performance implications. Based on our findings that performance impacts are prevalent and often unpredictable from change characteristics alone, proactive detection and analysis becomes essential. Automated profiling tools (such as Java Flight Recorder), static analysis for performance bottleneck identification, and machine learning-based anomaly detection can provide real-time insights into performance trends and deviations. Future research could focus on developing intelligent performance analysis systems that automatically classify performance changes as necessary trade-offs versus optimization opportunities, reducing the manual effort required for performance impact assessment while improving decision-making accuracy.

%% file: sections/threats.tex
\subsection{External Validity}

\ul{Number of Projects:} In this study, we analyzed 15 Java projects of varying size, complexity, and domain to ensure diversity in our dataset. Although this number aligns with multiple well-established studies~\citep{alcocer2019performance,alcocer2015tracking,sandoval2016learning,laaber2020dynamically,chen2019analyzing,reichelt2019peass}, the limited selection may have resulted in missing certain performance variation patterns that exist in other projects. Expanding the number of analyzed projects could potentially capture additional performance-related insights.

\ul{Focusing on Java Projects:} We selected Java for this study due to its widespread industry adoption and the availability of mature performance measurement tools. Given the unique characteristics of the Java Virtual Machine (JVM), the findings presented in this work may not directly generalize to other programming languages with different execution models. Additionally, continuous updates and optimizations introduced in newer Java versions may have influenced the observed performance behaviors, potentially leading to unintended variations across different versions.

\ul{Open-Source Projects Bias:} Our study exclusively focused on publicly available GitHub repositories with active development communities. While open-source projects often reflect real-world software development practices, they may not fully represent the performance-critical operations commonly found in proprietary or enterprise-level software. This limitation suggests that some performance variations relevant to closed-source systems might not have been captured in our analysis.

\subsection{Internal Validity}

\ul{Performance Measurement Accuracy:} To collect execution data of Java methods, we employed a lightweight instrumentation tool designed to minimize overhead while ensuring accurate measurement. Although the tool has undergone multiple validation tests to confirm its reliability in capturing execution times, the possibility of introducing minor inaccuracies due to measurement noise or external interference remains. Future work could explore complementary measurement techniques to further enhance accuracy.

\ul{Environmental Factors:} All benchmarking procedures were conducted on homogeneous machines under controlled conditions to minimize external interferences. However, factors such as background processes, system resource contention, or operating system variations could still have influenced the collected performance data. While we took precautions to mitigate these risks, completely eliminating environmental impact remains a challenge in performance benchmarking.

\subsection{Construct Validity}

\ul{Project Phase Division Methodology:} Our study segments project lifecycles into three equal-length time-based phases, which may not fully capture variations in commit activity across projects with different development rhythms. Some projects undergo intensive early development while others evolve steadily, potentially affecting the representativeness of lifecycle-based performance analysis. Alternative approaches such as commit-based phase division or hybrid strategies weighting time-based divisions with commit frequency could provide more adaptive segmentation aligned with actual development dynamics.

\ul{Effect Size Threshold Selection:} We used standard Cliff’s Delta thresholds to assess the magnitude of performance variations. While these thresholds have been widely adopted in prior research, the classification of effect sizes (i.e., small, medium, or large) may not always align perfectly with the actual performance impact in a given context. Alternative effect size definitions or adaptive threshold selection methods could improve the robustness of performance variation assessments.

\ul{Performance Metric Selection:} Execution time was chosen as the primary metric for assessing performance variations. However, additional performance indicators, such as CPU utilization, memory consumption, or I/O operations, could provide a more comprehensive view of performance changes. Future work could incorporate multiple performance metrics to enhance the accuracy of performance evaluations.

\ul{Code Change Classification and Labeling:} As outlined in Table~\ref{tab:code-change-categories}, we defined seven distinct code change categories based on prior research and extensive manual analysis of Java projects, with classification carried out by two authors with professional expertise in Java. While these categories effectively capture common types of modifications and achieved high inter-rater reliability, the categorization process involves inherent subjectivity where different reviewers may interpret certain modifications differently. Additionally, our taxonomy may not encompass all possible code changes across diverse software systems beyond Java projects. Future research could improve generalizability and consistency through expanded categorization schemas, automated classification models, or larger panels of expert reviewers.

\ul{Complexity Measure Validation:} A limitation of our study is that we did not empirically validate our weighted complexity measure against established baselines such as lines of changed code or change in cyclomatic complexity. While our approach is theoretically grounded in cognitive load research, future work should conduct a comparative analysis to quantify the predictive superiority of our composite measure over simpler metrics. Such validation would strengthen confidence in our complexity-based findings and provide empirical justification for the added methodological complexity of our weighted scoring approach.

\ul{Missing Data Points:} Throughout the process of building and benchmarking project commits, a substantial number of commits were excluded due to being unbuildable or lacking microbenchmark coverage for the modified methods. Consequently, significant performance variations (i.e., both improvements and regressions) may exist within these omitted data points. Future studies could investigate methods to recover missing data through dependency resolution or automated test generation techniques.

%% file: sections/conclusion.tex
This study contributes a large-scale empirical investigation of method-level performance evolution in {\numProjects} Java projects, providing a systematic analysis of {\numMethodChanges} method-level code changes. Our methodology establishes a replicable framework for analyzing performance evolution patterns using microbenchmarking, statistical analysis, and comprehensive categorization of code modifications that can be applied to future research in performance engineering.

The evidence presented challenges several common practices in software development. The widespread occurrence of performance impacts across all types of code changes questions the effectiveness of categorical risk assessment strategies, while the limited predictive power of traditional factors like developer experience and change complexity suggests that current performance management approaches may be insufficient. Our findings indicate that performance validation should be treated as a standard development practice rather than a specialized activity reserved for specific types of changes or project phases.

Our work provides practical guidance for improving performance management in software development. The empirical patterns we identified enable more informed decisions about testing resource allocation, team composition strategies, and domain-specific monitoring approaches. The comprehensive dataset we provide offers a valuable resource for developing automated performance analysis tools and prediction models. Looking forward, this research opens opportunities for investigating performance evolution in other programming languages, developing machine learning-based performance prediction systems, and creating intelligent development environments that proactively identify performance risks. By establishing empirical foundations for performance engineering decisions, this study contributes to more systematic and evidence-based approaches to software performance management.